\documentclass[12pt]{iopart}

\usepackage{iopams}
\usepackage{graphicx}
\usepackage[pdfpagelabels,colorlinks, citecolor=blue]{hyperref}
\begin{document}

\title[Probing of the interfacial Heisenberg and Dzyaloshinskii--Moriya exchange ... ]{Probing of the interfacial Heisenberg and Dzyaloshinskii--Moriya exchange interaction by magnon spectroscopy}

\author{Khalil Zakeri$^{1,2}$\footnote{Corresponding
author: khalil.zakeri@partner.kit.edu}}

\address{$^1$Heisenberg Spin-dynamics Group, Physikalisches Institut, Karlsruhe Institute of Technology,
Wolfgang-Gaede-Str. 1, D-76131 Karlsruhe, Germany}
\address{$^2$Max-Planck-Institut f\"ur Mikrostrukturphysik, Weinberg 2, D-06120 Halle, Germany}

\ead{khalil.zakeri@partner.kit.edu}

\begin{abstract} This Topical Review presents an overview of the recent experimental results on the quantitative determination of the magnetic exchange parameters in ultrathin magnetic films and multilayers, grown on different substrates. The experimental approaches to probe both the symmetric Heisenberg as well as the antisymmetric Dzyaloshinskii--Moriya exchange interaction in ultrathin magnetic films and at interfaces are discussed in detail. It is explained how the experimental spectrum of magnetic excitations can be used to quantify the strength of these interactions.
\end{abstract}

\vspace{2pc}
\noindent{\it Keywords}: Magnetic thin films and multilayers, Interface magnetic properties, Magnetic excitations, Magnetic exchange interaction, Heisenberg exchange interaction, Dzyaloshinskii--Moriya exchange interaction, Spin-polarized electrons, Spin-polarized electron energy loss spectroscopy.

\vspace{1cm}

\maketitle

\section{Introduction}

    Although the discovery of magnetism and magnetic materials dates back to earlier than 600 B.C., only after the development of the quantum theory in the twentieth century scientists have started to understand this fascinating effect. Today the research on magnetism is one of the major topics in condensed-matter physics. Of particular interest is the magnetism in low-dimensional solids. The motivation for investigation of magnetism in low-dimensions is twofold. Firstly, from the fundamental physics point of view, it is of great interest to see how the magnetic properties of a solid change when its dimensionality is reduced from a three dimensional bulk crystal to a two-, one- and zero-dimensional object. In many cases the change of the dimensionality leads to exotic effects, which are absent in the bulk magnets. Secondly, from a technological point of view, the integration of magnetic materials in new technologies requires these materials in the form of thin films, wires or dots. As a matter of fact low-dimensional magnetic structures lie in the central point of interest in magneto-electronics.

    In addition to the effects associated with the reduction of the system's dimensionality, there are other effects caused by the change of the chemical environment, which have a remarkable influence on the properties of low-dimensional magnetic solids. In practice both effects are used to tune the properties of low-dimensional magnets and to design structures with desired functionalities.

    The magnetic ground state of a given magnetic object is determined by different competing magnetic interactions. As a result, various types of magnetic structures, from simple collinear ferromagnetic (FM) and antiferromagnetic (AFM) to extremely complex noncollinear magnetic structures can be formed \cite{Sandratskii1998,Gao2007,Bergmann2008,Wulfhekel2010}. The formation of any kind of magnetic structure is the result of competing magnetic energies in the system, determining the ground state.  Therefore it is of utmost importance to figure out which interactions compete in a given object, forming that specific magnetic ground state. One of the most important magnetic interactions is the magnetic exchange interaction between neighboring atomic moments, known as Heisenberg exchange interaction (HEI). This interaction usually governs the magnetic energy of the system and determines the ground state. A quantitative measure of the strength of HEI in low-dimensional magnets is necessary to understand and predict the ground as well as the excited states of these systems. In particular, when a system exhibits an unusual pattern of exchange parameters, which can lead to very exotic magnetic states, this knowledge is of great importance.

    Another important magnetic interaction in low-dimensional magnets is the so-called antisymmetric Dzyaloshinskii--Moriya exchange interaction (DMI). It has first been shown by Dzyaloshinskii that in bulk systems with lack of the inversion symmetry there should exist a new type of interaction, different than HEI, which can explain the canted magnetic moments and hence the weak ferromagnetism, observed experimentally, in these materials \cite{Dzyaloshinskii1957}. It turned out that this interaction can be analytically derived when the spin--orbit part of the electronic Hamiltonian is included \cite{Moriya1960}. In the low-dimensional magnets, firstly the inversion symmetry is broken due to the presence of surfaces (interfaces), and secondly the spin--orbit coupling can be very strong, due to different chemical environment at the surfaces (interfaces). These two facts lead to a non-zero DMI, which can play a crucial role in the determination of the magnetic properties of low-dimensional solids \cite{Coffey1991, Zhao2003, DeRaedt2004, Sergienko2006, Bode2007, Ferriani2008, Heide2008, Heide2009, Hu2008, Katsnelson2010}.

    Here we review the recent experimental results on the quantitative determination of both HEI and DMI probed in ultrathin magnetic films and multilayers. Since layered magnetic structure are the building blocks of the present magneto-electronic technology, the focus is put on ultrathin magnetic films. The Topical Review is organized as following. In Sec. \ref{Sec:MagIntractions} the basic concepts of magnetic interactions are discussed. Different energy terms are introduced and their physical origin is explained. The mathematical description of magnetic excitations is presented in Sec. \ref{Sec:MagExcitations}. The experimental tools for probing these excitations are briefly introduced in Sec. \ref{Sec:ProbingMagExcitaions} with special attention to the electron scattering experiments, as those are the only experiments which can address magnetic excitations in low-dimensional structures. In Sec. \ref{Sec:ProbingHEI} the way of probing symmetric Heisenberg exchange interaction in layered magnetic structures is discussed in detail. It is discussed by presenting various examples that how the spectrum of magnetic excitations can be used to quantify the strength of the symmetric HEI. Section \ref{Sec:ProbingDMI} is dedicated to the antisymmetric DMI and how it can be quantified by looking at the dispersion relation of magnetic excitations. A brief summary is provided in Sec. \ref{Sec:Summary}.

    \section{Magnetic interactions} \label{Sec:MagIntractions}

     The important magnetic interactions in a magnetic solid are: (i) the symmetric Heisenberg exchange interaction, (ii) the antisymmetric Dzyaloshinskii--Moriya exchange interaction, and (iii) the long-range dipolar interactions. In order to introduce the magnetic interactions, we start with a general mathematical description of spin--spin interactions in a classical spin system. Note that this description is purely mathematical and does not identify the microscopic physical origin of magnetic interactions. We shall discuss the microscopic physical origin of each mathematical term in Secs. \ref{Sec:HEI} and \ref{Sec:DMI}, separably.

    We consider a lattice of spins $\mathbf{S}$  expanded in three dimensional space. In the Cartesian coordinate $\mathbf{S}$ can be expressed as a matrix with components $S_x$, $S_y$ and $S_z$. The most general form of the bilinear spin--spin interaction in such a lattice can be written in the following form

    \begin{eqnarray}
    \mathbf {I}=- \sum_{i, j} \mathbf{C} \mathbf{S}_{i} \mathbf{S}_{j} = - \sum_{i, j} \sum_{\alpha,\beta}C_{\alpha,\beta}S_{i,\alpha}S_{j,\beta},  \label{Eq:SpinHamiltonian1}
    \end{eqnarray}

    where $\mathbf{C}$ is the coupling matrix which shall couple $\mathbf{S}_{i}$ sitting on site $i$ to $ \mathbf{S}_{j}$ sitting on site $j$. The negative sign is used by convention to note that the energy of the system is minimized when all the spins are aligned parallel. In linear algebra any matrix can, in principle, be decomposed into a multiple of the identity matrix plus a traceless symmetric matrix plus an antisymmetric matrix. In the following we call the components of the symmetric matrix as $J_{ij}$ and the ones of the antisymmetric matrix as $D_{ij}$. One can therefore write $J_{ij}=-J_{ji}$ and $D_{ij}=-D_{ji}$. While the symmetric part of the coupling matrix can be written in the form of scaler products of spins  $\mathbf{S}_{i}$ and $\mathbf{S}_{j}$, the antisymmetric part can be described in terms of their vector products. The most simplest form of the spin Hamiltonian can therefore be written as:

    \begin{eqnarray}
    \mathcal{H}_{\rm Sym.} + \mathcal{H}_{\rm Antisym} = -\sum_{i\neq j}J_{ij}\mathbf{S_i}\cdot
    \mathbf{S_j}-\sum_{i\neq j}\mathbf{D}_{ij}\cdot\mathbf{S_i}\times
    \mathbf{S_j}.\label{Eq:SpinHamiltonianEx+DM}
    \end{eqnarray}

    In the following we discuss the physical origin of these two terms.

    \subsection{Symmetric Heisenberg exchange interaction} \label{Sec:HEI}
     The term introduced as the symmetric term in Eq. (\ref{Eq:SpinHamiltonianEx+DM}) is in fact the symmetric HEI and was first introduced by Werner Heisenberg at the beginning of the development of the modern theory of magnetism. Heisenberg could show that, in the case of electrons, the exchange interaction is a consequence of the Coulomb interaction between electrons and the Pauli exclusion principle and can be derived from the quantum mechanics.  Although there is no classical analogue to HEI, it may be considered as the overlap of the electronic wavefunctions of two (or more) identical electrons. HEI favors a collinear ground state. If the coupling constants $J_{ij}$ are positive (negative), a ferromagnetic (an antiferrmagnetic) ground state is favored leading to a parallel (an antiparallel) alignment of spins \cite{Heisenberg1928, Herring1966}.

    \subsection{Antisymmetric Dzyaloshinskii--Moriya exchange interaction} \label{Sec:DMI}

    The second term introduced in Eq. (\ref{Eq:SpinHamiltonianEx+DM}) is the antisymmetric DMI. This interaction was first proposed by Dzyaloshinskii in 1957. Based on symmetry arguments, he proposed this antisymmetric exchange term in order to explain the weak ferromagnetism observed in some materials like $\alpha-$Fe$_2$O$_3$ (Hematite) \cite{Dzyaloshinskii1957}. Almost at the same time it was shown by Moriya that, in principle, this interaction can be analytically derived by considering the relativistic spin--orbit correction in the Hamiltonian of interacting electrons \cite{Moriya1960}. DMI is essential to understand many physical properties of different systems for example spin-glasses \cite{Fert1980}, cuprates \cite{Coffey1991}, molecular magnets \cite{Zhao2003, DeRaedt2004} and multiferroics \cite{Sergienko2006,Hu2008}.

    Unlike HEI, DMI favors a noncollinear magnetic ground state. For a system of interacting electrons one can show that DMI is a consequence of the relativistic spin--orbit coupling and absence of the inversion symmetry in the system. For the systems with inversion symmetry this term vanishes. In the case of ultrathin magnetic films the presence of the surfaces and interfaces breaks the inversion symmetry. Hence one would expect that DMI is active in such structures and may lead to very exotic ground states \cite{Bode2007, Ferriani2008, Heide2008, Heide2009, Hu2008}.

\subsection{Dipolar interaction and magnetic anisotropy}
    In principle a spin sitting on a lattice site can be coupled to the magnetic stray field generated by the other spins located in longer distances, describing a long-range spin--spin interaction. Although this term is similar to HEI, as far as the symmetry is concerned, it has a very different nature than HEI. The dipolar interaction is given by
    \begin{eqnarray}
    \mathcal{H}_{\rm
    Dip}=-\sum_{ij}\frac{(g\mu_B)^2}{r_{ij}^3}\Bigg[\mathbf{S_i}\cdot\mathbf{S_j}-3(\mathbf{\hat{r}_{ij}}\cdot\mathbf{S_i})
    (\mathbf{\hat{r}_{ij}}\cdot\mathbf{S_j})\Bigg],\label{Eq:SpinHamiltonianDipolar}
    \end{eqnarray}
    where $\mathbf{r_{ij}}$ represents the displacement vector of spins $\mathbf{S_i}$ and $\mathbf{S_j}$, located at $\mathbf{r_{i}}$ and $\mathbf{r_{j}}$, respectively. This interaction is much weaker than the HEI and is usually responsible for the demagnetizing field and formation of ferromagnetic domains in ferromagnets.

    The classification of spin--spin interactions described above is based on two important assumptions (i) the spins (magnetic moments) are considered as rigid entities, and (ii) only the bilinear terms are considered.

    In the case of $3d$ ferromagnets, however, the magnetism is attributed to the itinerant electrons. One may raise the question: How the spin Hamiltonian described above can be applied to such systems. Although in itinerant ferromagnets the magnetism is caused by itinerant electrons \cite{Moriya1984}, one may consider that the electrons are partially localized on atoms. This allows one to define the atomic magnetic moment. By associating a magnetic moment (spin) to each atomic site one can use the above mentioned treatment. Such a treatment may not provide the full description of the underlying physics. It, however, provides a way of estimating the strength of the magnetic interactions among different magnetic moments in the system.

    An expansion of Eq. (\ref{Eq:SpinHamiltonian1}) beyond the bilinear interaction of spins is certainly possible. This would lead to additional terms which are proportional to $S^4$,  $S^6$ and so on. One can show that those type of interactions are of symmetric type. However, they are much weaker that the HEI. In some cases it is necessary to consider those terms to correctly describe the magnetic properties of the system \cite{Bergmann2007,Ferriani2008,Heinze2011}. Since we do not treat them in the present review, we do not discuss them further.

\section{Magnetic excitations} \label{Sec:MagExcitations}

    \subsection{Magnetic excitations within the localized moment picture} \label{Sec:MagExLocalPic}
    The spin Hamiltonian discussed above describes both the ground as well as the excited state of a spin system. The bosonic excitations describing the collective excitations of the system are magnons. Mathematically they are the eigenstates of the spin Hamiltonian. They may be described as the wave-like excitations caused by the precessional spins about the equilibrium direction. The hallmark of a magnon is its total angular momentum, which is $1\hbar$. This unique property of magnons makes the identification of magnons possible, among all other excitations in solids.

    In order to derive an equation for the magnon dispersion relation one needs to calculate the eigenvalues of the spin Hamiltonian. This can either be done by using classical dynamics (see for example Ref. \cite{Keffer1953}) or by the so-called Holstein-Primakoff approach (see for example Refs. \cite{Holstein1940, Dyson1956, VanKranendonk1958}) for any system of interest. Since the symmetry of the system and the arrangement of spins is very important in the spin Hamiltonian given in Eq. (\ref{Eq:SpinHamiltonianEx+DM}), it is not possible to derive a general equation for the dispersion relation. For each system of interest one has to solve the equations separately, considering the geometry of the system.  In the case of a cubic lattice with only one magnetic atom in the unit cell the dispersion relation of the symmetric HEI can be written as:

    \begin{eqnarray}
    \varepsilon &=& 2kJ_{n}S
    \left[1-\frac{1}{k}\sum_{\mathbf{R}} \cos \left(\mathbf{Q} \cdot \mathbf{R}\right)\right] \nonumber\\
    &+& 2hJ_{2n}S\left[1-\frac{1}{h}\sum_{\mathbf{R^{\prime}}} \cos \left(\mathbf{Q} \cdot \mathbf{R^{\prime}}\right)\right] \nonumber\\
    &+& 2lJ_{3n}S\left[1-\frac{1}{l}\sum_{\mathbf{R^{\prime\prime}}} \cos \left(\mathbf{Q} \cdot \mathbf{R^{\prime\prime}}\right)\right] + ...,
    \label{Eq:DispersionRelation}
    \end{eqnarray}
    where $\varepsilon$ is the magnon energy, $k$, $h$ and $l$ are the numbers of nearest, second nearest and third nearest neighbors, respectively, $S$ represents the magnitude of spin,  $J_n$, $J_{2n}$ and $J_{3n}$ are the symmetric Heisenberg exchange coupling constants between the nearest, second nearest and third nearest neighbors, $\mathbf{Q}$ is the magnon wave vector and $\mathbf{R}$, $\mathbf{R^{\prime}}$ and $\mathbf{R^{\prime\prime}}$ represent the position vector of the respective neighbors. For the limit of small wave vectors the magnon energies can be approximated by $\varepsilon= D_{HEI}Q^2$, where $D_{HEI}$ is the well-known magnon stiffness coefficient and is given in the units of meV\AA$^{2}$ (or THz\AA$^{2}$).

    Equation (\ref{Eq:DispersionRelation}) indicates that the symmetric HEI leads to a fully symmetric magnon dispersion relation. This means that by reversing the sign (the direction) of $\mathbf{Q}$ the magnon energy remains unchanged i.e., $\varepsilon(Q)=\varepsilon(-Q)$. In addition, it is rather straightforward to see from Eq. (\ref{Eq:DispersionRelation}) that reducing the dimensionality of the system can lead to substantial changes in the magnon dispersion relation. For example, if the dimensionality of a 3D bulk ferromagnet is reduced to a 2D system, one would expect a drastic change in the magnon dispersion relation, due to the fact that the number of neighbors is changed.

    In the case of layered structures composed of $m$ atomic layers one expects $m$  magnon modes. The dispersion relation of the symmetric HEI can be calculated by starting from Eq. (\ref{Eq:SpinHamiltonian1}) and finding the solution of the following matrix equation:

    \begin{eqnarray}
    \varepsilon \left(\matrix{A_1 \cr
                                            .\cr
                                             .\cr
                                             .\cr
                                          A_m\cr}\right) =  \left(\matrix{2S\sum^m_j J_{ij}\left[A_1-A_j \exp \left(i \mathbf{Q} \cdot \mathbf{R_{ij}}\right)\right] \cr
                                            .\cr
                                             .\cr
                                             .\cr
                                          2S\sum^m_j J_{ij}\left[A_m-A_j \exp \left(i \mathbf{Q} \cdot \mathbf{R_{ij}}\right)\right]\cr}\right); \hspace{5mm} \mathbf{R}=\mathbf{R_i}-\mathbf{R_i}.
    \label{Eq:Amplitude}\end{eqnarray}

    Equation (\ref{Eq:Amplitude}) implies that for a slab composed of $m$ atomic layers one should observe $m$ different magnon modes. There will be two magnon modes which are lower in energy with respect to the others. Those modes are formed due to the lower coordination number of spins in the top and bottom layers of the slab. At $Q=0$ these two modes are separated by $\varepsilon (Q=2\pi/m)$ in energy. They degenerate in energy at high wave vectors, close to the zone boundary, if the same values of exchange constants are considered at the top and bottom layers. The quantities $A_i$ represent the amplitude of the eigenvectors. It is apparent from Eq. (\ref{Eq:Amplitude}) that this quantity depends on the exchange constants as well. For a given magnon mode $A_i$ varies from one layer to another. Therefore this quantity might be considered as the contribution of each layer to that particular magnon mode. For example the two lowest-energy magnon modes have the largest amplitude of the eigenvectors in the top and bottom layers. This means that these modes are localized at these two layers. If these two layers are magnetically identical it is not possible to distinguish between them. However, if the layers are magnetically different (the exchange constants in the layers are different) the degeneracy of these two low energy magnon mods breaks and one can associate a mode to each layer.

    The lowest in energy magnon mode, which satisfies the Goldstone criteria $\varepsilon(Q=0)=0$, is the so-called \emph{``acoustic mode"}. The higher energy magnon modes are the so-called \emph{``optical modes"}, which possess a finite energy at $Q=0$, in analogy to the phonon modes in solids.

    Now we consider the DMI term in the analysis of magnon dispersion relation. The energy associated with this interaction can be expressed as \cite{Udvardi2003, Udvardi2009}:

    \begin{eqnarray}
    \varepsilon_{\rm{DMI}} &=&  c \sin^2 \theta\Bigg[\mathbf{D}_1 \cdot \mathbf{\hat{e}} \sum_{\mathbf{R}} \sin \left(\mathbf{Q} \cdot \mathbf{R}\right)\\ &+& \mathbf{D}_2 \cdot \mathbf{\hat{e}} \sum_{\mathbf{R^{\prime}}} \sin \left(\mathbf{Q} \cdot \mathbf{R^{\prime}}\right)+\mathbf{D}_3 \cdot \mathbf{\hat{e}} \sum_{\mathbf{R^{\prime\prime}}} \sin \left(\mathbf{Q} \cdot \mathbf{R^{\prime\prime}}\right) +...\Bigg].
    \label{Eq:DispersionRelationDM}
    \end{eqnarray}

    Here $c$ is the so-called chirality rotation index and is defined with respect to the direction of the easy axis (being $+1$ for right rotating sense and $-1$ for the left rotating one), $\theta$ denotes the relative angle between spins and the direction of the easy axis $\mathbf{\hat{e}}$, $\mathbf{D}_1$, $\mathbf{D}_2$ and $\mathbf{D}_3$ are the DM vectors of the nearest, second nearest and third nearest neighbors, respectively,  $\mathbf{R}$, $\mathbf{R^{\prime}}$ and $\mathbf{R^{\prime\prime}}$ represent the position vector of the respective neighbors. Equation (\ref{Eq:DispersionRelationDM}) implies that only the components of the DM vector parallel to the easy axis contribute to the magnon energies. It is apparent that for the limit of small wave vectors this term can be approximated by $D_{\rm{DMI}}Q$, where $D_{\rm{DMI}} = c \sin^2 \theta \left((\mathbf{D}_1\cdot \mathbf{\hat{e}}) \sum_{\mathbf{R}} R_{\parallel}+ (\mathbf{D}_2\cdot \mathbf{\hat{e}}) \sum_{\mathbf{R^{\prime}}} R_{\parallel}^{\prime}+ (\mathbf{D}_3\cdot \mathbf{\hat{e}}) \sum_{\mathbf{R^{\prime\prime}}} R_{\parallel}^{\prime\prime} + ... \right)$ may be called the DMI induced magnon stiffness constant and is given in the units of meV\AA$^{1}$ (or THz\AA$^{1}$). Note that this quantity is somewhat different than the well-known magnon stiffness coefficient associated with the HEI.

    For a thin ferromagnetic film in the presence of a large spin--orbit coupling and in the absence of the inversion symmetry both the symmetric exchange interaction and the antisymmetric DM interaction are active. In this case the resulting magnon dispersion relation will include the term introduced in Eq. (\ref{Eq:DispersionRelation}) and also the one introduced in Eq. (\ref{Eq:DispersionRelationDM}). As a result the dispersion relation will be asymmetric with respect to $Q$, meaning that $\varepsilon(Q)$ is no longer equal to $\varepsilon(-Q)$. We will come to this point in Sec. \ref{Sec:ProbingDMI}.

    As discussed above the dynamical response of the magnetic system includes all the interactions within the system and hence probing the dispersion relation would lead to an unambiguous determination of both HEI and DMI. In particular in the case of low-dimensional ferromagnets, where it is expected that these interactions are totally different than the bulk, probing of these quantities is of utmost importance for understanding their properties. We will show in Secs. \ref{Sec:ProbingHEI} and \ref{Sec:ProbingDMI} by showing various examples how this can be done.

    It is worth mentioning that a precise determination of these fundamental interactions is only possible when the dispersion relation is measured over a wide range of $Q$. This means that only the experimental techniques enabling the measurement of magnetic excitations over a large fraction of the Brillouin zone are appropriate for such studies.

    \subsection{Magnetic excitations within the band formalism}
    Based on the band theory of magnetism, in a ferromagnetic metal the degeneracy of the electronic bands is lifted
    due to the electron-electron interactions \cite{Stoner1936, Stoner1938, Slater1937, VanKranendonk1958, Kubler2000} This degeneracy breaking leads to a splitting of the electronic bands for majority and minority electrons by the so called ``exchange energy".  In the case that the Fermi-level is located between the lower (majority) and higher energy  (minority) bands, it results a net spin polarization, which explains the ferromagnetic state of the system. This spin-split bands across the Fermi-level can, in principle, lead to the possibility of a single-electron excitations across the Fermi-level. For instance an electron from an occupied majority band jumps to an unoccupied state in the minority band after reversing its spin. This process leads to the formation of an electron-hole pair (Stoner pair) with the angular momentum of $1\hbar$. It is important to notice that Stoner excitations can occur over a relative large area of the Brillouin zone. From the band picture of magnetism the magnons can be described as a coherent superposition of Stoner pairs. The wavefunction of a magnon is a linear superposition of the wavefunctions of electron and hole states forming that particular magnon state.
    This main difference between localized moments ferromagnets and itinerant electron ferromagnets leads to the fact that in the latter case the collective magnon modes dissipate into single particle Stoner excitations. As a result magnons in itinerant ferromagnets possess a very short lifetime. The decay of magnons into Stoner excitations which is usually called Landau damping depends strongly on the available Stoner states in the same energy region as magnons. Note that in ultrathin ferromagnets grown on a substrate the location and the shape of the Stoner density of states depends also on the degree and the type of electronic hybridizations between the film and substrate \cite{Buczek2011a, Buczek2011b, Qin2015}.

    As mentioned above, although in itinerant ferromagnets the magnetism is caused by the itinerant electrons, one may consider that the electrons are partially localized on atomic sites. This allows one to associate a magnetic moment to each atomic site. The wave-like excitations of these magnetic moments that propagate through the solid show collective properties. Again in this case it is possible to describe the magnons by a Heisenberg type of Hamiltonian. In such a case the pairwise interactions of magnetic moments are regarded as effective interactions which couples the moments.

    \section{Probing magnetic excitations} \label{Sec:ProbingMagExcitaions}

    Magnetic excitations can be investigated by various scattering techniques e.g. inelastic neutron, x-ray and electron scattering techniques. In the following we shall briefly mention the physical principles behind each of these techniques.
    \subsection{Probing magnetic excitations by neutrons}
    When neutrons are scattered from a magnetic material they can excite magnetic excitations. In the inelastic neutron scattering (INS) a neutron beam with a given energy is scattered from the sample and thereby the magnetic excitations are excited. The scattered neutron carries all the information regarding the wave vector and energy of the created magnon. The spin of the incoming neutron interacts with the magnetic moment of the unit cell and excites magnons in the material. The physical interaction which leads to the magnon excitation is of pure dipolar nature (spin--spin interaction i.e., the dipolar interaction between the neutron's spin and the magnetic moment of the unit cell). Since this interaction is rather weak, INS does not allow probing magnons in low-dimensional magnets. The technique has been extremely successful for probing the dispersion relation of magnetic excitations in various bulk magnetic materials \cite{Shirane2002, Chatterji2005, Willis2009}.
    \subsection{Probing magnetic excitations by photons}
    Magnons can be excited by inelastic scattering of light from a magnetic sample. In Brillouin light scattering (BLS) experiments, the magnons are excited and probed by photons. The frequency shift of the incoming photon is measured after the scattering. The coupling mechanism behind the excitation of magnons is the modulation of the dielectric constant of the material with light via the magneto-optical effects. When a photon is incident to a magnetic material, due to the magneto-optical effects, a phase grating is generated by a magnon. This phase grating propagates in the material with the phase velocity of the magnon. Consequently, when the light is scattered from this phase grating, its frequency is Doppler-shifted by the frequency of the magnon \cite{Grunberg1985,Hillebrands1989,Demokritov2001,Hillebrands2003}.

    BLS can be used to investigate the magnons with small wave vectors near the center of the Brillouin zone. In order to probe the magnetic excitations with large wave vector, one may use the resonant processes. This is the basis of resonant inelastic x-ray scattering (RIXS) \cite{Ament2011}. As the direct interaction between the photon and the electorns' spin via the magnetic part of the electromagnetic field is extremely small, a direct photon-electron interaction is not efficient. Therefore, one may take advantage of resonant process of photon absorption. In RIXS an x-ray photon is inelastically scattered from the sample and the energy, and polarization change of the scattered photon is measured. In RIXS the energy of the incident photon is tuned such that it matches to one of the x-ray transition edges of the material. The resonance process has the following advantages. First, it can greatly enhance the cross section of the inelastic light scattering. Second, one can excite and probe the magnetic excitations of a complex multi-element crystal in an element selective manner. Third, since the resonant process involves the core electrons, one takes the advantage of the large spin-orbit coupling of core electrons and efficiently transfer the photon angular momentum to the spin system. It has been shown that this technique can be applied to complex bulk samples, mainly magnetic oxides \cite{Ament2011}.

    \subsection{Probing magnetic excitations by spin-polarized electrons} \label{Sec:SPEELS}
    Magnetic excitations can be excited very efficiently by means of spin-polarized electrons \cite{Qin2015, Plihal1999,Vollmer2003, Ibach2003, Etzkorn2004, Prokop2009, Zakeri2013, Zakeri2014}. In the spin polarized electron energy loss spectroscopy (SPEELS) a low-energy electron beam with a well-defined energy and spin polarization is scattered from the sample surface and the energy loss (gain) of the scattered electrons is measured. In a ferromagnetic sample with a defined magnetization, magnons can be excited (annihilated) by incidence of minority (majority) electrons via a spin-flip process, as a magnon possesses a total angular momentum of $1\hbar$. Analysis of the spectra recorded for different spin polarizations of the incoming electron beam (parallel and antiparallel to the sample magnetization) leads to the identification of magnons. At the first glance one may think that the SPEELS technique provides the same information as INS. This is in fact not true, as the fundamental principles of these two techniques are different. The excitation process in SPEELS is mediated by the quantum mechanical exchange interaction, which is of a pure Coulomb nature and not on the magnetic dipole interaction as in INS. Since the interaction of electrons with the matter is very strong (much stronger than the neutron-matter interaction), this technique is extremely sensitive and can be used to measure the magnons in an ultrathin film with the ultimate thickness i.e. one (or even sub) atomic layer \cite{Prokop2009, Zakeri2014a, Qin2015}. This makes the SPEELS technique a unique method for probing collective excitations at surfaces and in low-dimensional solids.

    Similarly, inelastic scanning tunneling spectroscopy (ISTS) has also been successfully used to investigate the collective spin excitations in thin films with a high energy and spatial-resolution \cite{Balashov2006, Balashov2008, Gao2008, Balashov2014}. Unlike SPEELS, the technique cannot be used to excite and investigate these excitations in a wave vector selective manner. While SPEELS enables one to measure the dispersion relation of collective excitations with a high momentum and energy resolution, ISTS provides the excitation spectra with a high energy and spatial resolution.

    \section{Probing the Heisenberg exchange interaction in ultrathin ferromagnetic films} \label{Sec:ProbingHEI}

    \subsection{Fe monolayer on W(110)} \label{Sec:FeMLW}

    The most straightforward way to probe the interfacial exchange interaction is to probe the magnon dispersion relation of a magnetic monolayer. The SPEELS technique has opened such a possibility. Fe monolayer on W(110) has been chosen as a prototype system. The advantage of Fe(110) monolayer on W(110) is that it grows psudomorphically with rather large smooth traces of one atomic layer height. In addition, the system exhibits a high thermodynamic and chemical stability. This means that the Fe and W atoms at the interface do not mix with each other, forming a complex interface. As the interface is very sharp, the system can be regarded as a model system. The Curie temperature of Fe monolayer on W(110) is about 223 K \cite{Elmers1994, Elmers1995, Elmers1996}.

    \begin{figure}[t!]
    \begin{center}
    \includegraphics[width=0.95\textwidth]{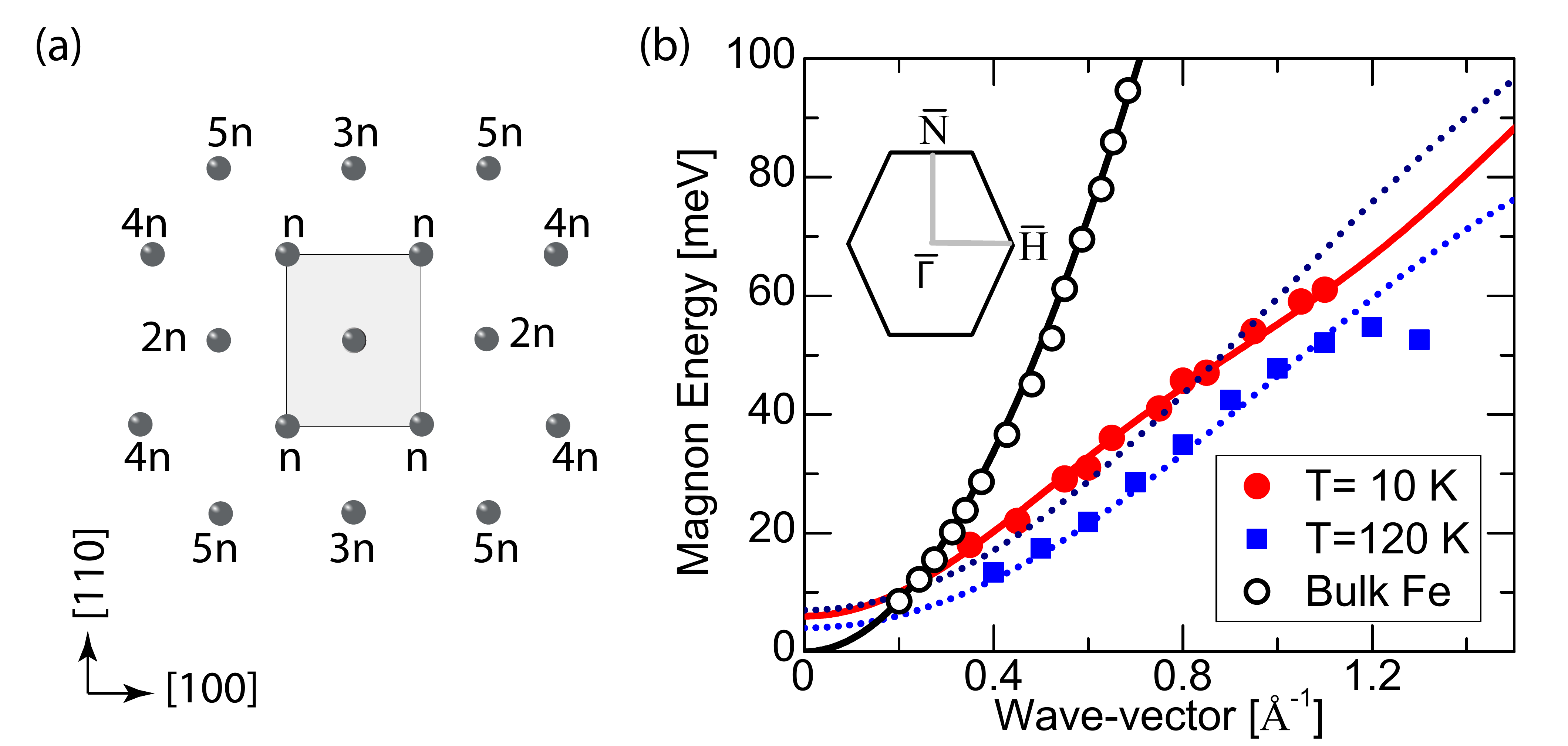}
    \caption{\label{Fig:DispersionML}  (a) The real-space arrangement of Fe atoms in an Fe(110) monolayer. The unit cell is depicted with the rectangle. The corresponding surface Brillouin zone is shown in the inset of (b). (b) The magnon dispersion relation of Fe monolayer on W(110). The experimental data are shown by solid symbols while the fits based on the Heisenberg model are shown by solid (considering up to the fifth nearest neighbors) and dotted (considering only the nearest neighbors) curves. The fitting parameters are listed in Tab. \ref{Tab:Monolayer}. The experimental data are from Ref. \cite{Prokop2009} (T=120 K) and Ref. \cite{Zakeri2014a} (T=10 K). The bulk dispersion relation measured by INS by Lynn and Mook \cite{Lynn1981} is shown for a comparison.}
    \end{center}
    \end{figure}

    The magnon dispersion relation of a ferromagnetic monolayer has first been measured at 120 K \cite{Prokop2009} and later on at 10 K \cite{Zakeri2014a}. The results of those experiments are summarized in Fig. \ref{Fig:DispersionML} and are compared to the room temperature magnon dispersion relation of bulk Fe, measured by INS \cite{Lynn1981}. At the first glance one observes that the magnon dispersion relation in the Fe monolayer is very much different than the one of the bulk Fe. The ``soft" magnons in the case of Fe monolayer on W(110) may be caused by two different reasons. (i) The first reason might be the reduction of the number of nearest neighbors. As it is apparent from Eq. (\ref{Eq:SpinHamiltonianEx+DM}) the magnon energies depend strongly on the symmetry of the system and also on the number of the neighboring spins. (ii) The second reason might be that the HEI at the interface is substantivally different than the one of the bulk Fe. It will be shown in the following that this is indeed the case.

    Fitting the INS data measured by Lynn and Mook \cite{Lynn1981} with the bulk dispersion curve results in a value of about $J_{n}S$ = 13.2 meV if one sets the higher order terms to zero and  $J_{n}S$ = 8.9 meV if one assumes $J_{2n}=0.6J_nS$ and $J_{3n}=J_{4n}S = ...=0$. Similarly, by starting with Eq. (\ref{Eq:DispersionRelation}) one can derive the following  equation for the magnon dispersion relation of the Fe(110) monolayer:

    \begin{eqnarray}
    \varepsilon &=&  8J_{n}S \left[1- \cos \left(\frac{Q_{\parallel}a }{2}\right)\right]  + 4J_{2n}S \Bigg[1- \cos\left(Q_{\parallel}a\right)\Bigg] \nonumber \\&+& 8J_{4n}S \left[1- \cos \left(\frac{3}{2}Q_{\parallel}a \right)\right] + 8J_{5n}S \Bigg[1- \cos \left(Q_{\parallel}a \right)\Bigg] \nonumber \\
    &+& 4J_{6n}S \Bigg[1- \cos \left(2Q_{\parallel}a \right)\Bigg]\label{Eq:DispersionFeML}
    \end{eqnarray}

  Using the values obtained for bulk Fe one fails to explain the experimental magnon dispersion relation [see Fig. \ref{Fig:DispersionML} (b)].

    \begin{table}
    \caption{\label{Tab:Monolayer} The Heisenberg exchange coupling constants in Fe monolayer on W(110). The experimental results are obtained by fitting the magnon dispersion relation with an extended Heisenberg model. The results of ab initio calculations by means of different approaches are shown for a comparison. The values are given in meV. To better compare the experimental results to the ab initio results the value of $S$ is set to 2.2. The definition of abbreviations used in the table are as follows LKGM: Liechtenstein-Katsnelson-Gubanov method, FMA: frozen magnon approximation, DLM: disordered local moments, SKKR: screened Korringa-Kohn-Rostoker method, TB: tight-binding scheme,  RS-LMTO-ASA: real-space linear-muffin-tin-orbital approach implemented in the atomic sphere approximation.}
    \begin{center}\begin{tabular*}{0.8\textwidth}{@{}l*{15}{@{\extracolsep{0pt plus12pt}}l}}
    \br
    & $J_n$ & $J_{2n}$ & $J_{3n}$ &$J_{4n}$ & $J_{5n}$\\
    \mr
    Experiment at 120 K \cite{Prokop2009} & $12.1\pm1$ & set to $0$ & set to $0$ & set to $0$ & set to $0$\\
    Experiment at 10 K \cite{Prokop2009} & $14.3\pm1$ &set to $0$ &set to $0$ & set to $0$ & set to $0$\\
    Experiment at 10 K \cite{Zakeri2014a} & $13.2\pm1$ & $-9.9\pm1$ & --- & $1.3 \pm 1$ & $4.1\pm1$\\
    LKGM \cite{Bergman2010} & $14$ & $-11.5$ & $3.5$ & $4$ & $0.2$\\
    FMA \cite{Bergman2010} & $12$ & $-9$ & $2$ & $2.5$ & $0.5$\\
    DLM \cite{Bergman2010} & $17.5$ & $-9.5$ & $2.5$ &$2$ & $0.7$\\
    SKKR \cite{Udvardi2009} & $10.8$ & $-3.3$ &$3.6$ &$4.6$ & ---\\
    TB \cite{Muniz2003,Costa2003,Muniz2008,Costa2008} & $42.5$ &$3.7$ & $0.5$ & $-0.9$ & $-0.5$\\
    TB  + KKR-GF \cite{Costa2010} & $28.7$ & $-7.9$ & $0.3$ & --- & ---\\
    TB + RS-LMTO-ASA \cite{Costa2010} & $11.2$ & $-7.3$ & $0.2$ & --- & ---\\
    \br
    \end{tabular*}
    \end{center}
    \end{table}
    Heisenberg exchange coupling constants can be calculated from first principles. There are a few numerical schemas for calculating the Heisenberg exchange parameters. In all calculations one starts with the ab initio calculations of the electronic structures of the system. The pairwise HEI between the magnetic atoms are then calculated based on adiabatic approximation (for a summery of different schemes for calculating the magnon dispersion relation and more detail on technical implementations see for example section 5.1 of Ref. \cite{Zakeri2014} and references therein). Such calculations for ferromagnetic films on a nonmagnetic substrate have revealed that the Heisenberg exchange coupling constants at interfaces are very much different than the corresponding bulk values. A summary of the calculated Heisenberg exchange coupling constants for Fe monolayer grown on W(110) is provided in Tab. \ref{Tab:Monolayer}. The origin of this difference lies in both the reduction of the dimensionality of the system and also the effects associated with the presence of the W(110) substrate. It has been shown that the main source of this difference is the strong electronic hybridizations at the interface of Fe with W which hinders the overlap of the electronic wavefunctions of neighboring magnetic atoms and leads to a weaker HEI \cite{Prokop2009,Bergman2010,Chuang2012}. It has been observed in most of the calculations that in contrast to the bulk Fe the exchange constants in the Fe monolayer on W(110) are rather long range.  This means that the interactions beyond nearest and next nearest neighbors are important to be considered when describing the magnon dispersion relation. The exchange interaction can in some cases be of antiferromagnetic character (negative exchange constant). Most of the calculations predict that the next nearest neighbor coupling constant is of antiferromagnetic character ($J_{2n}<0$).

     One may try to adopt the calculated dispersion relation to the experimental one [see Fig. \ref{Fig:DispersionML} (b)]. Assuming a negative value for $J_{2n}$, one can fit  the experimental data rather good with Eq. (\ref{Eq:DispersionFeML}). The results of such analysis are listed in Tab. \ref{Tab:Monolayer} and are compared to the results of first principles calculations.

    The negative exchange coupling constant shall, in principle, have a direct impact on the magnon dispersion relation. In the case of the Fe(110) monolayer, due to the symmetry of the system, in the equation derived for the magnon dispersion relation along the $\bar{\Gamma}$--$\bar{\rm{H}}$  some terms do not appear and some other cancel each other \footnote{For instance the term including $J_{3n}$ does not appear in the magnon dispersion relation due to symmetry of the bcc(110) surface. In addition, the terms including $J_{2n}$ and $J_{5n}$ are almost identical, meaning that a negative $J_{2n}$ is easily compensated by a positive $J_{5n}$. Since $J_{5n}$  describes the coupling between the atoms located at a larger distance compared to $J_{2n}$, its magnitude should, in principle, be smaller. However the number of the fifth neighbors are two times larger than the second nearest neighbors (see Fig. \ref{Fig:DispersionML} (a))}. Therefore it is not easy to directly see the consequence of the antiferromagnetic HEI on the dispersion relation. We will discuss in Sec. \ref{Sec:DirectEvidAFM} that in the case of the bcc(001) surface the influence of the antiferromagnetic HEI on the magnon dispersion relation can be easily seen by looking at the magnon dispersion relation, in particular at the high symmetry points.

    \subsection{Fe bilayer on W(110)}

    \begin{figure}[b!]
    \begin{center}
    \includegraphics[width=0.5\textwidth]{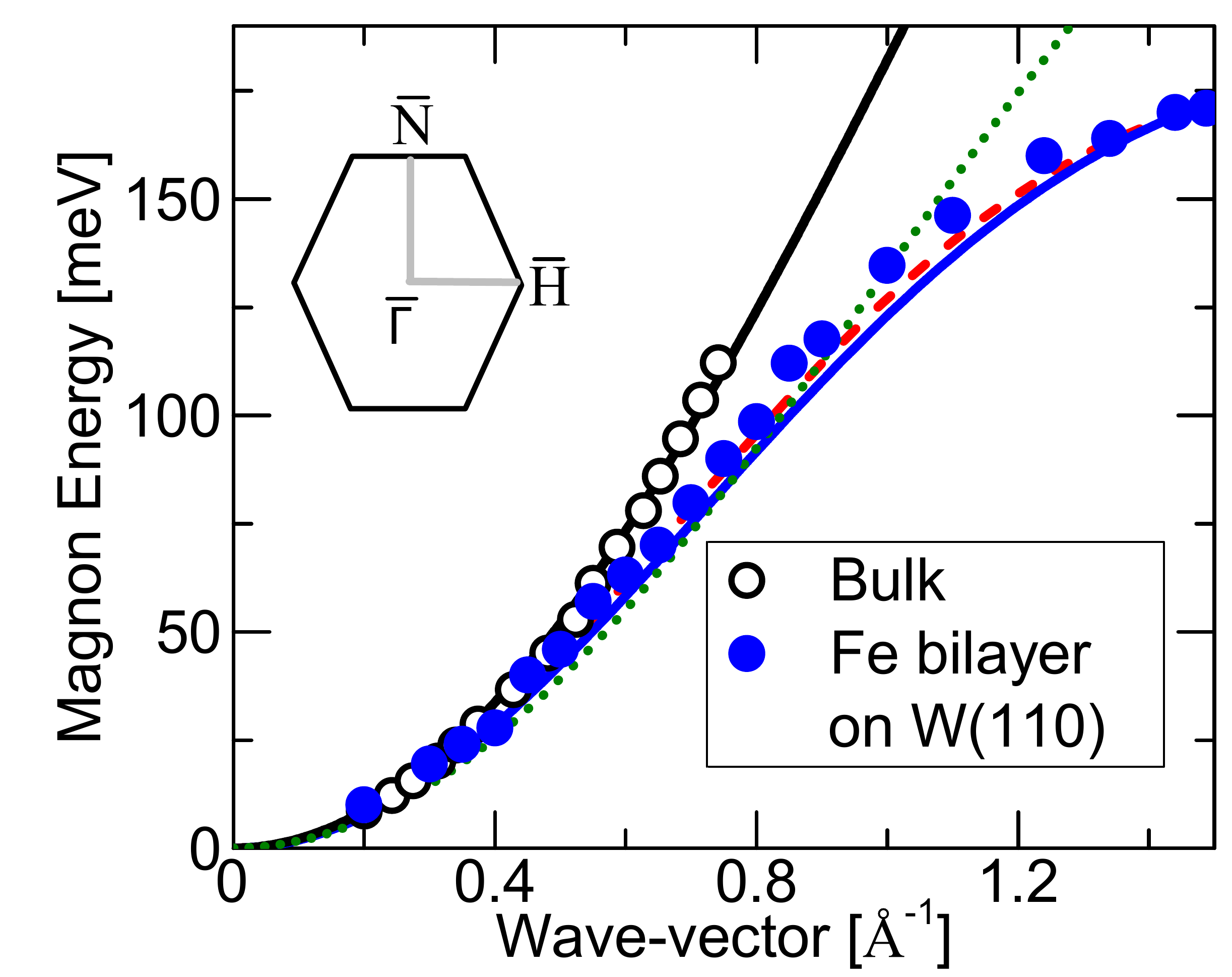}
    \caption{\label{Fig:DispersionBI}  The magnon dispersion relation of the Fe bilayer on W(110). The bulk dispersion relation is shown for a comparison. The dotted curve is a fit based on the Heisenberg model (Eq. (\ref{Eq:DispersionFeBIAcostic})) including only an isotropic nearest neighbor coupling ($\widetilde{J_{n}^{\parallel}}=\widetilde{J_{n}^{\perp}}$  and $\widetilde{J_{Nn}^{\parallel}}=0$ ). The dashed curve represents the case where an isotropic nearest neighbor coupling is taken into account and the higher order terms are also included ($\widetilde{J_{n}^{\parallel}}=\widetilde{J_{n}^{\perp}}$  and $\widetilde{J_{Nn}^{\parallel}}=0.6 \widetilde{J_{n}^{\parallel}}$ ). The solid curve represents the case in which the inter- and intralayer exchange constants are not the same and the higher order interactions are also taken into account ($\widetilde{J_{n}^{\parallel}}\neq\widetilde{J_{n}^{\perp}}$  and $\widetilde{J_{Nn}^{\parallel}}=0.6 \widetilde{J_{n}^{\parallel}}$ ). The resulted exchange coupling constats are listed in Tab. \ref{Tab:Bilayer}.}
    \end{center}
    \end{figure}

    \begin{table}[h]
    \begin{center}
    \caption{\label{Tab:Bilayer} The Heisenberg exchange coupling constants in the Fe bilayer on W(110) obtained by fitting the experimental data with an extended Heisenberg model. The results of calculations by means of different approaches are shown for a comparison. The values are given in meV. To better compare the experimental results to the results of calculations the value of $S$ is set to 2.2. In the definition of abbreviations used in the table are as follows LKGM: Liechtenstein-Katsnelson-Gubanov method, MFT: magnetic force theorem.}
    \begin{tabular*}{0.75\textwidth}{@{}l*{15}{@{\extracolsep{0pt plus12pt}}l}}
    \br
    & $\widetilde{J_{n}^{\parallel}}$ & $\widetilde{J_{n}^{\perp}}$ & $\widetilde{J_{Nn}^{\parallel}}$ \\
    \mr
    Experiment at 300 K \cite{Tang2007} & $24.2\pm2$ & $24.2\pm2$ & $0$ \\
    Experiment at 300 K \cite{Tang2007} & $16.5\pm1$ & $16.5\pm1$ & $0.6\left(\widetilde{J_{n}^{\parallel}}\right)$ \\
    Experiment at 300 K \cite{Tang2007} & $13.5\pm1$ & $24.2\pm2$ & $0.6\left(\widetilde{J_{n}^{\parallel}}\right)$ \\
    Ab initio calculation Ref.  \cite{Chuang2012} & $16$ & $28$ & ---\\
    Ab initio calculation Ref. \cite{Bergqvist2013} & $13.5$ & $14.6$ & --- & \\

    \br
    \end{tabular*}
    \end{center}
    \end{table}

    The Heisenberg exchange couplings discussed in Sec. \ref{Sec:FeMLW} are confined in the plane of the Fe monolayer. It is of great interest to see how the pattern of the Heisenberg exchange coupling constants changes when a second layer of Fe is deposited on top. First principles calculations predict that the pattern of the exchange interaction changes drastically when the second Fe layer is added on top. Analyzing the experimental magnon dispersion relation may help to prove this prediction. Fitting the measured magnon dispersion relation of the Fe bilayer on W(110) with the Heisenberg model is a difficult task, as one has to deal with too many unknown parameters. For each atomic layer there are $n$ unknown exchange parameters. In addition the coupling of layers has to be described by additional parameters. In our analysis we use the following simplifications. The nearest neighbors coupling in each layer is put into an \emph{``effective intralayer coupling"} $\widetilde{J^{\parallel}_{n}}$. Similarly, the term describing the coupling between the two layers is considered as an \emph{``effective interlayer coupling"} $\widetilde{J^{\perp}_{n}}$. The higher order interactions are considered as an effective interaction, which may in turn include several terms. Starting with Eq. (\ref{Eq:Amplitude}) one can derive the following equations for the magnon dispersion relation of the Fe(110) bilayer (a slab including two atomic layers of Fe(110)):

    \begin{eqnarray}
    \varepsilon & = &  \left(8\widetilde{J_{n}^{\parallel}}S + 4\widetilde{J_{n}^{\perp}}S\right) \left[1- \cos \left(\frac{Qa }{2}\right)\right] + 4\widetilde{J_{Nn}^{\parallel}}S \Bigg[1- \cos\left(Qa\right)\Bigg].
    \label{Eq:DispersionFeBIAcostic}
    \end{eqnarray}

    \begin{eqnarray}
    \varepsilon & = &  8\widetilde{J_{n}^{\parallel}}S  \left[1- \cos \left(\frac{Qa }{2}\right)\right] + 4\widetilde{J_{n}^{\perp}}S  \left[1+ \cos \left(\frac{Qa }{2}\right)\right] \nonumber\\ & + & 8\widetilde{J_{Nn}^{\perp}}S + 4\widetilde{J_{Nn}^{\parallel}}S \Bigg[1- \cos\left(Qa\right)\Bigg].
    \label{Eq:DispersionFeBIOptic}
    \end{eqnarray}

    The superscript $\parallel$ ($\perp$) in Eqs. (\ref{Eq:DispersionFeBIAcostic}) and (\ref{Eq:DispersionFeBIOptic}) indicates the coupling in the same atomic plane (between different atomic planes) which is usually referred to as intralayer (interlayer) coupling. Equation (\ref{Eq:DispersionFeBIAcostic}) satisfies the Goldstone theorem and is the acoustic magnon mode of the system. Results based on the fitting of the measured magnon dispersion relation with Eq. (\ref{Eq:DispersionFeBIAcostic}) are presented in Tab. \ref{Tab:Bilayer}. As it is expected, the value of $\widetilde{J_{n}^{\parallel}}$ is larger than the nearest neighbor interaction $J_n$ of the Fe monolayer. This is because this value is the average value of $J_n$ of the surface and interface layers. The value of $\widetilde{J_{n}^{\perp}}$ is larger than the one of $\widetilde{J_{n}^{\parallel}}$, because the interatomic distances between atoms from two layers is smaller than the one of the atoms form the same layer. Smaller interatomic spacing means a larger overlap of the electronic wavefunctions and consequently a stronger HEI.

    It is important to mention that in the analysis presented in Fig. \ref{Fig:DispersionBI} DMI is not taken into account. It will be shown in Sec. \ref{Sec:ProbingDMI} that DMI has its consequence on the breaking of the symmetry of the magnon dispersion relation (see Sec. \ref{Sec:ProbingDMI}).

    \subsection{Fe and Fe/Co multilayers on Ir(001)} \label{Sec:Fe_Ir}

    As discussed in Sec. \ref{Sec:MagExLocalPic}, based on Eq. (\ref{Eq:Amplitude}) one can show that if the exchange parameters in different layers are different, the amplitudes $A_i$ of each magnon mode will be different in each layer.This is a very common situation in layered structures due to the fact that the atomic environment of the surface and interface atoms is drastically different from the one of the atoms located in the inner part of the film. In addition, due to the epitaxial growth the distances between layers can vary when moving from the interface towards the surface. Consequently different exchange parameters are expected for atoms sitting in different layers (the exchange parameters become layer dependent). In particular, for the atoms sitting in the interface layer the exchange parameters can be very much different than those of the atoms sitting in the other layers. As a result the pattern of the magnetic exchange parameters across the film can be very complicated. In many combinations of $3d$ ferromagnetic films with nonmagnetic substrates HEI in the layer adjacent to the substrate is weaker than the one in the surface layer. This is a consequence of the electronic hybridizations of the $3d$ electronic states of the first atomic layer of the ferromagnetic film with the substrate states. This fact has an impact on the magnon dispersion relation. In such a case the lowest-energy magnon mode of the system has the largest spectral weight when it is projected into the interface layer. One may say that the lowest-energy acoustic magnon mode is mainly localized at the interface. In a classical picture, this means that the precessional amplitude of the moments forming this particular magnon mode has its maximum value when it is projected into the interface layer. Hence by probing this lowest-energy magnon mode one can quantify the Heisenberg exchange parameters in the interface layer.
    Such an approach has been used to probe the exchange parameters in epitaxial Fe films and Fe/Co multilayers grown on the Ir(001) and Rh(001) substrates \cite{Zakeri2013c}.

 \begin{figure}[b!]
\begin{center}
\includegraphics[width=0.98\textwidth]{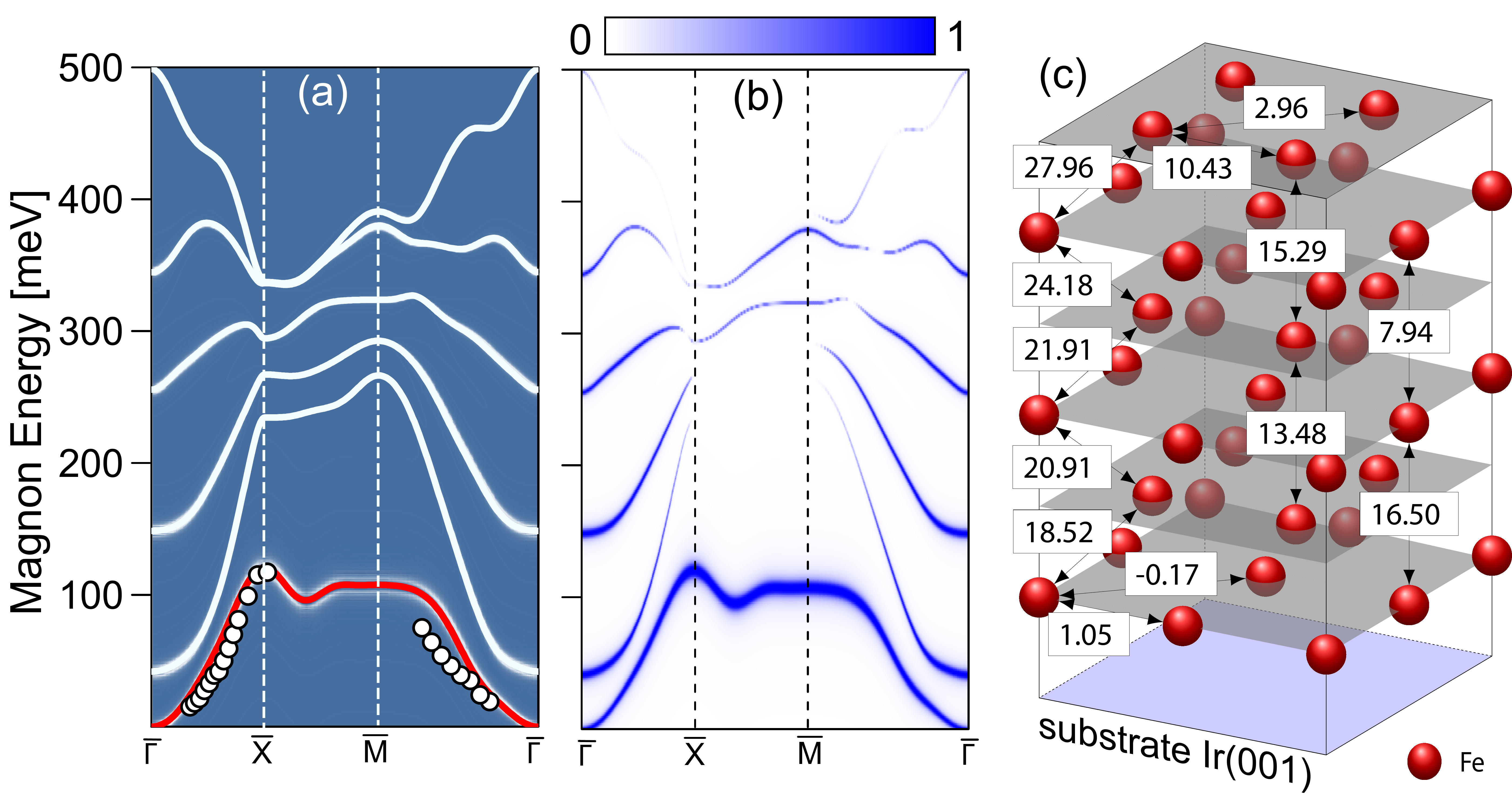}
\caption{\label{Fig:DispersionFe_Ir}  The magnon dispersion relation of a 6 ML Fe film on Ir(001). (a) The calculated susceptibility Bloch's spectral function of all Fe layers are plotted together with the experimental data. The white color indicates the places where the magnon band exist. The eigenvalues of the Heisenberg Hamiltonian are represented by solid lines. The lowest energy magnon band is highlighted by the red color. (b) The susceptibility Bloch's spectral function projected into the interface Fe layer. The largest spectral weight is observed for the lowest-energy magnon mode when the spectral function is projected into the interface layer, meaning that this mode is formed due to the weak exchange interaction in the interface layer. (c) The calculated interatomic exchange parameters resulted from the ab initio calculations. All the values are given in the units of meV.  Adopted from Zakeri et al. \cite{Zakeri2013c}. Copyright (2013) by Macmillan Publishers Ltd: Nature Nanotechnology.}
\end{center}
\end{figure}

The dispersion relation of the lowest-energy magnon mode measured by SPEELS on a 6 ML Fe film on Ir(001) is presented in Fig. \ref{Fig:DispersionFe_Ir} (a) by open circles. Comparing to the results of Fe bilayer on W(110), presented in Fig. \ref{Fig:DispersionBI},  one recognizes that the magnon energies are lower in the present case. In particular the magnon energies along the $\bar{\Gamma}$--$\bar{\rm{M}}$ direction are about half of the values measured in the case of Fe bilayer on W(110). One may try to model the system by using a Heisenberg model. However such an attempt does not lead to reliable results, as one has to deal with a large number of unknown parameters. In this case ab initio calculations are very useful for quantifying the values of the exchange parameters.  A comparison of the measured magnon dispersion relation to the results obtained by means of ab initio calculations leads to the determination of the exchange coupling constants. In Fig. \ref{Fig:DispersionFe_Ir} (a) all the modes predicted by the adiabatic magnon calculations are shown. The lowest energy mode (the acoustic mode) fits very well to the experimental data. In order to figure out whether this mode is formed due to the weak HEI at the interface one way would be to calculate the Bloch's spectral function. The amplitude of the dynamic susceptibility is directly proportional to the amplitude of the eigenvectors ($A_i$ in Eq. (\ref{Eq:Amplitude})). When the susceptibility Bloch's spectral function is projected into the interface layer it can be regarded as the contribution of the interface layer to the magnon modes of the system. Figure \ref{Fig:DispersionFe_Ir} (b) shows the projected susceptibility Bloch's spectral function projected into the interface layer.  At the zone center $\bar{\Gamma}$  point ($Q=0$) the interface layer contributes identically to all magnon modes of the system.  The contribution of the interface layer to the lowest energy mode becomes larger as one goes from the $\bar{\Gamma}$ point towards the zone boundaries. At the $\bar{\rm{X}}$-point and along the $\bar{\rm{X}}$--$\bar{\rm{M}}$ direction the interface layer contributes only to the lowest energy mode, meaning that this part of the magnon spectrum is an interface part. If one performs similar calculations for a free-standing slab composed of 6 layers of Fe(001) one would observe again 6 magnon modes. The two lowest energy modes will degenerate in energy at the high symmetry $\bar{\rm{X}}$ and $\bar{\rm{M}}$ points. In addition, since in this case the surface and interface layers are equivalent the spectral weight of the two low-energy modes projected into the surface layer will be identical to the one projected into the interface layer. In the case of real systems (ferromagnetic film grown on the a nonmagnetic substrate) the presence of the substrate breaks the degeneracy of the two low-energy magnon modes and leads to the formation of a mode which is lower in energy. This lowest-energy mode originates from the interface layer, where the exchange coupling constants are smaller. The exchange coupling constants predicted by the ab initio calculations are shown in Fig. \ref{Fig:DispersionFe_Ir} (c). As it is apparent from Fig. \ref{Fig:DispersionFe_Ir} (c), the pattern of the exchange interaction in the Fe/Ir(001) system is very complicated. The exchange parameters are strongly layer dependent. While the interlayer coupling for all layers is strongly ferromagnetic, the intralayer exchange coupling at the interface is very weak. The values of the exchange constants at the interface are very small and in some cases negative, meaning that the Fe monolayer grown on Ir(001) has a tendency to be ordered antiferromagnetically.

\begin{figure}[h!]
\begin{center}
\includegraphics[width=0.98\textwidth]{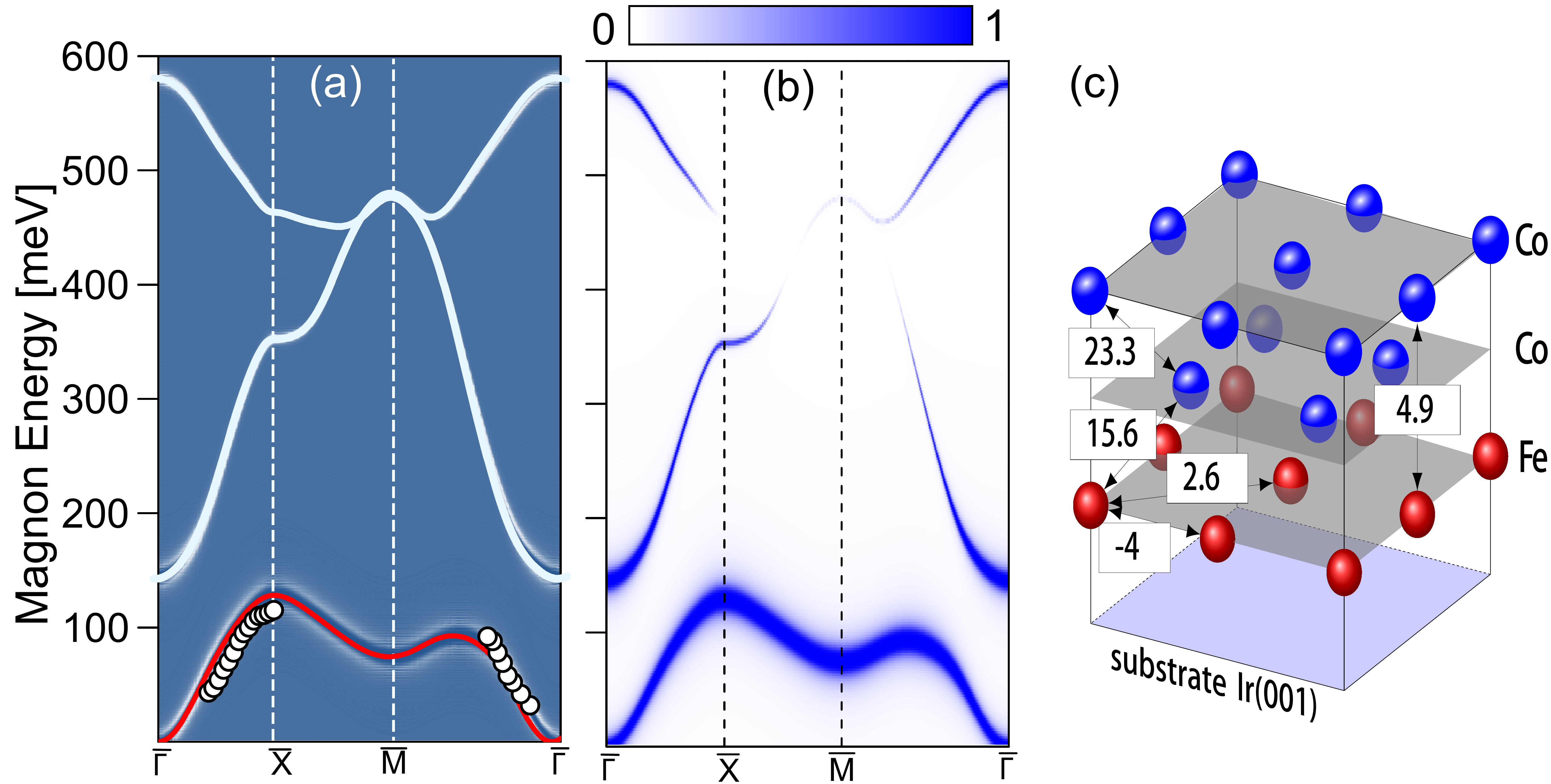}
\caption{\label{Fig:DispersionCoFe_Ir}  The magnon dispersion relation of 2 ML Co on 1 ML Fe on Ir(001). (a) The calculated susceptibility Bloch's spectral function of all layers are plotted together with the experimental data. The white color represents the peak position of the spectral function. The magnon bands (the eigenvalues of the Heisenberg Hamiltonian) are represented by solid lines. The lowest energy magnon band is highlighted by the red color. (b) The susceptibility Bloch's spectral function projected into the interface Fe layer. The largest spectral weight is observed for the lowest-energy magnon mode when the spectral function is projected into the interface layer, meaning that this mode is formed due to the weak exchange interaction in the interface Fe layer. (c) The calculated interatomic exchange
parameters resulted from the ab initio calculations. All the values are given in meV.  Adopted from Zakeri et al. \cite{Zakeri2013c}. Copyright (2013) by Macmillan Publishers Ltd: Nature Nanotechnology.}
\end{center}
\end{figure}

In order to ensure that it is the interface mode which has been observed in the experiment one may grow an Fe monolayer on the Ir(001) surface and cover it by Co layers. It is well known that the magnon energies in the case of Co thin films are at higher with respect to Fe layers \cite{Vollmer2003, Costa2004, Etzkorn2005, Taroni2011,
Rajeswari2012, Ibach2014, Rajeswari2014, Rajeswari2013a, Michel2015}. Hence if a similar low energy magnon mode is observed in the experiment, one can make sure that this mode is in fact  formed at the interface Fe layer. The experimental magnon dispersion relation measured for a multilayer composed of one monolayer of Fe buried under 2 and 3 layers of Co is presented in Fig. \ref{Fig:DispersionCoFe_Ir}. The magnon energies are very similar to the case of 6 ML Fe on Ir(001). This is a clear hint that the low energy magnon mode observed in the experiment is indeed an interface mode. In the same way as discussed above the magnon dispersion relation is calculated by means of ab initio theory. The calculated magnon modes and the Bloch's spectral function projected into the interface layer are presented in Fig. \ref{Fig:DispersionCoFe_Ir} (a) and (b), respectively.
The values of the exchange coupling constants are depicted in Fig.  \ref{Fig:DispersionCoFe_Ir} (c). The calculated Bloch's spectral function indicates that the low-energy acoustic magnon mode is mainly an interface mode. Again in this case the exchange coupling constants at the interface Fe layer are very small. Remarkably, the nearest neighbor intralayer exchange coupling constant is negative, meaning that this interaction is of antiferromagnetic character. We will discuss this point in greater detail in Sec. \ref{Sec:FeCo_Ir}.

    \begin{figure}[t!]
    \begin{center}
    \includegraphics[width=0.65\textwidth]{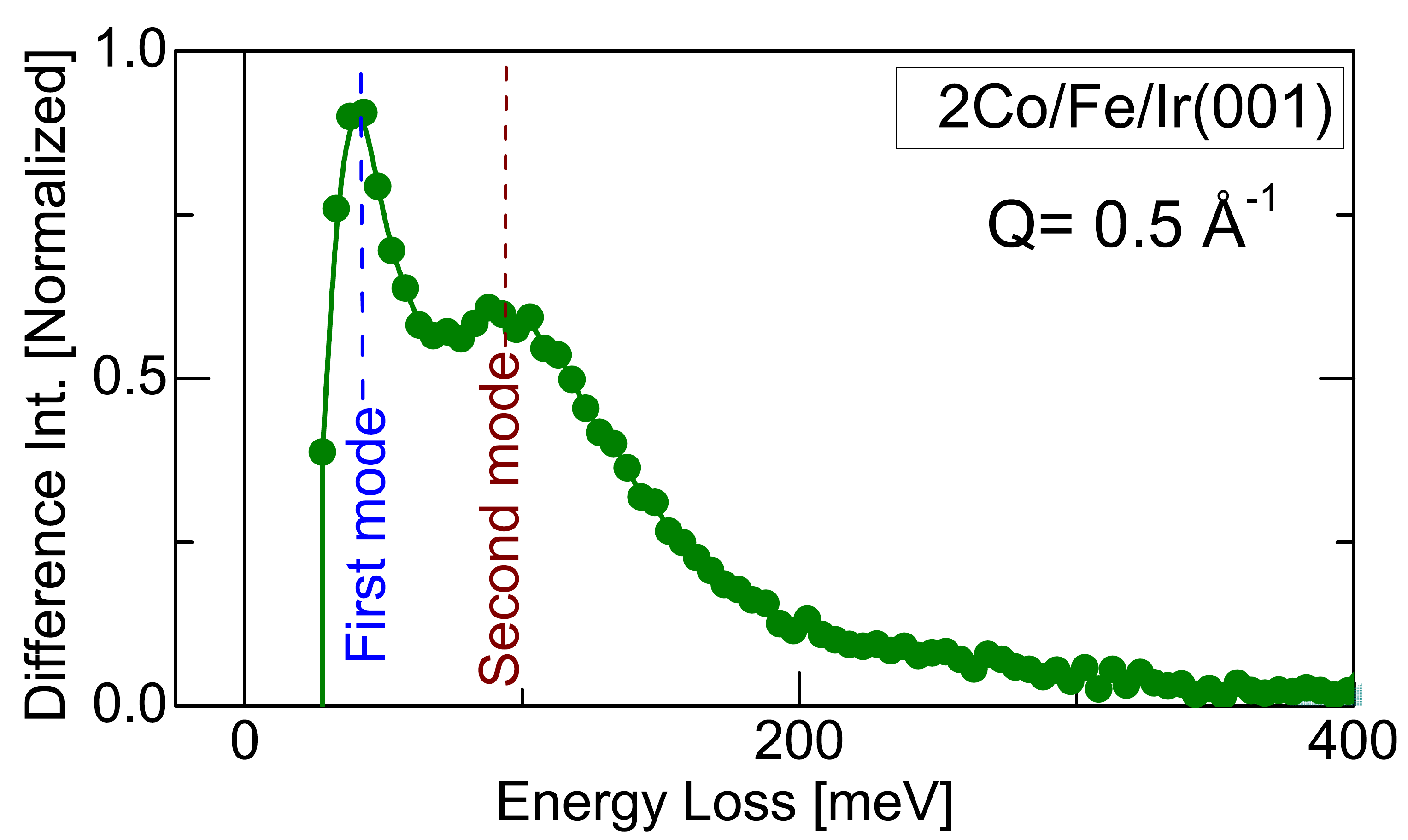}
    \caption{\label{Fig:SpectraCoFeIr}  Normalized difference spectra recorded on a Co/Co/Fe trilayer structure
    grown on Ir(001). One can clearly observe at least two magnon modes. The modes are well-separated. The presence of the low-energy magnon mode is due to the weak exchange interaction in the interface Fe layer. The data are recorded at a wave vector of $Q=0.5$ \AA$^{-1}$.}
    \end{center}
    \end{figure}

It is worth mentioning that for both 6 ML Fe/Ir(001) and 2MLCo/1MLFe/Ir(001) systems more than one magnon mode has been detected in the experiment. The attention is paid to the lowest-energy mode, as it is relevant for further considerations. In the case of 2MLCo/1MLFe/Ir(001) the second magnon mode is very obvious and can be assigned to the surface Co layer. An example of the SPEELS difference energy-loss-spectra recorded on this system is presented in Fig. \ref{Fig:SpectraCoFeIr}. In the SPEELS experiments the spectra are usually recorded for two possible spin orientations of the incoming electron beam \cite{Plihal1999,Vollmer2003,Tang2007,Prokop2009,Zakeri2014}. The conservation of the angular momentum during the scattering prohibits magnon excitations for incoming electrons with a spin polarization antiparallel to the sample magnetization. Only electrons having spin parallel to the sample magnetization can excite magnons (see Sec. \ref{Sec:SPEELS}). This fact leads to peaks in the difference spectra, defined as $I_{\rm{Diff.}}=I_{\downarrow}-I_{\uparrow}$, at particular loss energies where the magnons are excited. The spectrum shown in Fig. \ref{Fig:SpectraCoFeIr} is recorded at a wave vector of $Q=0.5$ \AA$^{-1}$ in the midway of $\bar{\Gamma}$--$\bar{\rm X}$. As one may notice from the data shown in Fig. \ref{Fig:SpectraCoFeIr} the energy of the mode associated with the surface Co layer (the second mode shown in Fig. \ref{Fig:DispersionCoFe_Ir}) is overestimated in the adiabatic calculations shown in Fig. \ref{Fig:DispersionCoFe_Ir}. This is a general observation. In the case of Co layers the magnon energies of all magnon modes measured in the scattering experiments are lower in energy compared to the results of the adiabatic calculations (as well as the dynamical calculations) \cite{Vollmer2003, Costa2004, Etzkorn2005, Taroni2011,
Rajeswari2012, Ibach2014, Rajeswari2014, Rajeswari2013a, Michel2015}. In the experiments energies of up to 250 meV are observed for the acoustic magnon mode, whereas the calculations predict energies of up to 500-600 meV. One of the possible explanations might be that the underlying electronic structures of Co films are not correctly described by density functional theory (DFT), as the correlation effects may not be fully captured within this framework (see Sec.  \ref{Sec:Co_Cu} for more discussions on the exchange parameters in Co films). In the case of Fe films, however, ab initio DFT calculations provide a very good estimation of the magnetic exchange parameters. Furthermore, the dynamical calculations based on the same Kohn-Sham potentials predict perfectly the lifetime of excitations in such films \cite{Buczek2011a, Zakeri2013, Qin2015}.

Note that this problem that DFT based calculations do not precisely predict the exchange parameters in Co layers does not alter the validity of the method explained above (the values might be overestimated by 15$\%$  \cite{Rajeswari2014}). As discussed in Sec. \ref{Sec:MagExcitations} even with a very simple model one can show that the spectral weight of the lowest-energy magnon mode of a film composed of a few atomic layers has the largest amplitude in the layer, in which the exchange parameters are smaller. The fact that the exchange interaction in the Fe layer is much weaker than the one in Co layers on top is justified by both theory and experiment. The dispersion relation of the acoustic magnon mode of the 2MLCo/1MLFe/Ir(001) system shows very low energies, only up to 110 meV at the $\bar{\rm{X}}$--point and about 92 meV at the $\bar{\rm{M}}$--point (see Fig. \ref{Fig:DispersionCoFe_Ir}). Such a low-energy magnon mode is an indication of a weak exchange interaction in the buried Fe layer.  As the exchange interaction in this layer is very weak, the spectral weight of this lowest-energy mode should have its maximum when it is projected into this interface Fe layer. The pattern shown as Bloch's spectral function of magnons in Figs. \ref{Fig:DispersionCoFe_Ir} (a) and (b) remains nearly unchanged when the values of exchange parameter in Co layers given in Fig. \ref{Fig:DispersionCoFe_Ir} (c) are reduced by 85\%. The position of the lowest-energy magnon mode remains also unchanged. Probing of this mode provides all the necessary information regarding the magnetic exchange parameters in the Fe layer.

\subsection{Co films on Cu(001) and W(110)} \label{Sec:Co_Cu}

The first attempt of estimating the magnetic exchange interaction in Co films was done when the magnon dispersion relation of fcc and hcp Co films could be measured by SPEELS \cite{Vollmer2003, Vollmer2004a, Vollmer2004b, Vollmer2004c, Etzkorn2004}.  It was realized that magnon energies measured in the experiment are smaller than the values calculated by theory. However, the measured magnon dispersion relation of both fcc and hcp films could be well described by the simple nearest neighbor Heisenberg model, assuming an isotropic HEI. This assumption has resulted in a value of $J_n=15\pm0.1$ meV, very close to the value of bulk Co measured earlier by INS experiments. The magnon mode observed in the experiment was attributed to the surface Co layer and it was discussed that \emph{``it cannot be ruled out that the peak contains an unresolved optical mode of significant lower intensity, since the peak shape is slightly asymmetric"} \cite{Vollmer2003}. More recent experiments by means of EELS could confirm that at low wave vector region the two lowest-energy magnon modes are well-separated and can be resolved \cite{Rajeswari2012, Ibach2014, Rajeswari2014, Rajeswari2013a, Michel2015}.  The spectra in the region of low wave vectors were measured with a higher energy resolution, while at higher wave vectors the same resolution as in Ref. \cite{Vollmer2003} was used to record the data. At the high wave vectors, energy of the two lowest-energy magnon modes is very close to each other so that they cannot be separated. Attempts have been made to separate the different magnon modes of Co films using a simple classical models based on the finite mean free path of electrons \cite{Ibach2014}. However, a precise separation of different magnon modes requires an unambiguous description of the scattering intensity. A complete consideration of the inelastic electron scattering processes, leading to the creation of magnetic excitations, together with the corresponding scattering matrix elements are needed in order to describe the intensity of the SPEELS spectra. Unfortunately, such a theory does not exist at the present time.

Experiments on Cu and Ni covered Co films have shown magnon energies which are just slightly smaller than the case without capping layer \cite{Rajeswari2013a}. Simple analysis based on the nearest neighbor Heisenberg model have revealed that the value of $J_nS$ is reduced by 10 and 15 $\%$, when an 8 ML Co film is covered by Cu and Ni, respectively. The reduction of $J_nS$  may be due to the weaker exchange interaction and/or the reduced magnetic moment. The surface magnetic moment is usually larger than the interface one \cite{Bluegel1996}. It is also larger than the one of the atoms located in the inner part of the film. For instance the Co magnetic moment at the fcc(100) or hcp(0001) surfaces is  by 15\% and 7\% larger than the bulk moment. It reaches the bulk value already in the first, latest in the second layer below the surface \cite{Bluegel1996}. The surface magnetic moment is usually suppressed when a capping layer is added. In addition to that, effects due to the intermixing caused by the formation of the alloys when the capping material is introduced cannot be excluded. Therefore, it is not straightforward to attribute the reduction of $J_nS$ to the weakening of the Co--Co exchange interaction at the interface only. If one assumes that the surface magnetic moment is unchanged while capping the Co film (which is not a correct assumption) one would realize that the exchange interaction of Co atoms at the interfaces are weaker than the one at the surface. The fact that the exchange parameters at Co/Cu interface is just slightly smaller than the ones at the Co/vacumme interface has been observed in different calculations \cite{Costa2004, Buczek2010a, Buczek2011a, Zakeri2013c, Bergqvist2013, Etz2015}. For the case of Fe, however, it has been observed that the exchange parameters at the surface and interface can be entirely different. The surface and interface exchange parameters can even exhibit different signs (the origin of this effect is discussed in Secs. \ref{Sec:Fe_Rh}, \ref{Sec:FeCo_Ir}, \ref{Sec:DirectEvidAFM}).

\begin{figure}[t!]
    \begin{center}
    \includegraphics[width=0.65\textwidth]{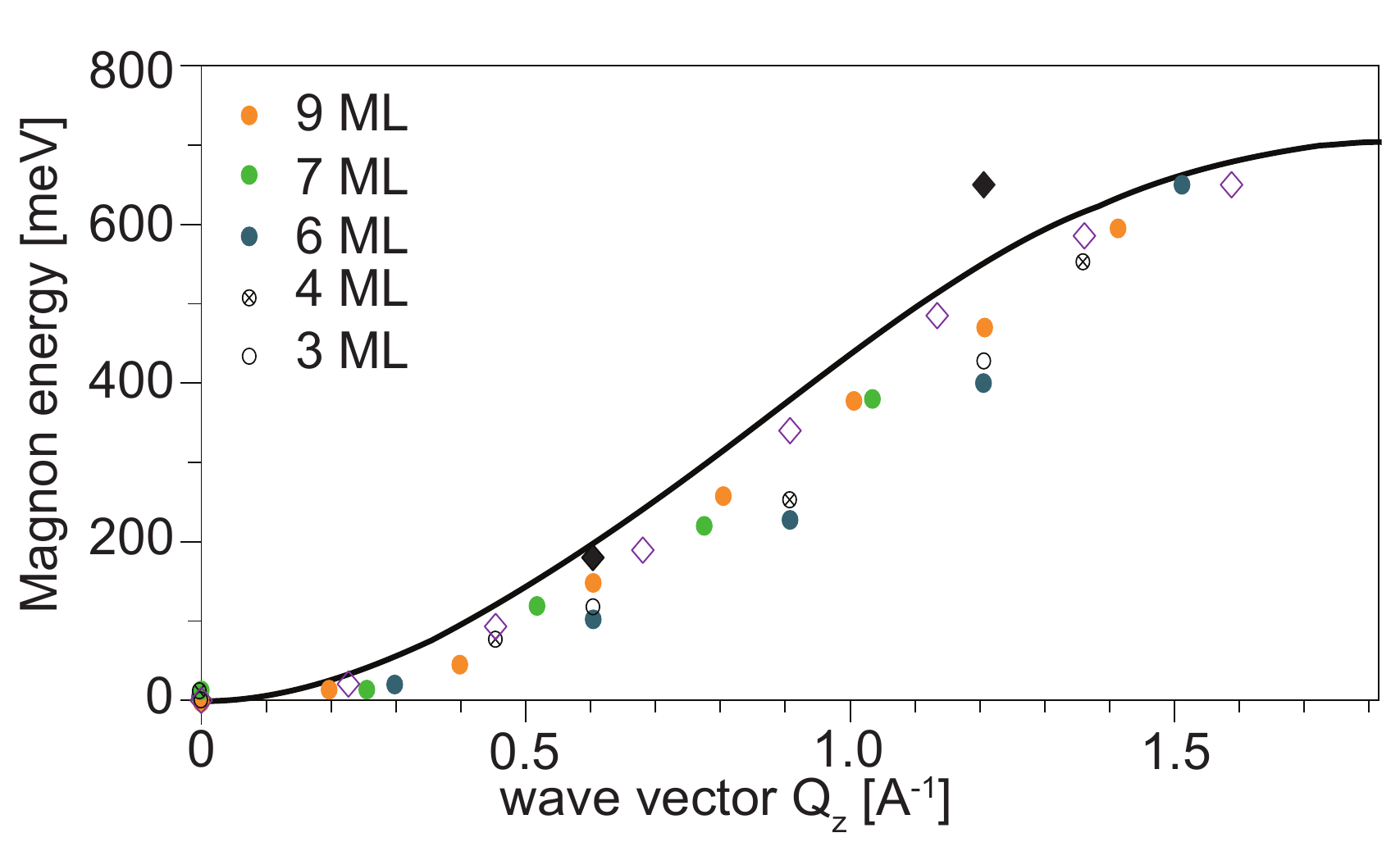}
    \caption{\label{Fig:DispersionCoCu}  The magnon dispersion in Co films grown on Cu(001) probed by ISTS (open and solid circles). The dispersion relation is constructed by matching peaks (observed in positive bias voltages) and dips (observed in the negative bias voltages) of the inelastic tunneling spectra. The solid curve is the dispersion relation of bulk fcc Co calculated by Pajda et al. \cite{Pajda2001}. The solid and open diamonds correspond to energies of the magnon bands at the zone center for 3 and 8 ML Co on Cu(001), calculated by Bergqvist et al. \cite{Bergqvist2013}. Adopted from Balashov et  al. \cite{Balashov2014}.  Copyright 2014 by IOP Publishing.}
    \end{center}
    \end{figure}

In Ref. \cite{Rajeswari2014} it has been suggested that the layer-resolved exchange coupling in Co films can be tested by analysing the dispersion relation of the two low magnon modes of the system at small wave vectors. An EELS spectrometer has been used to probe the dispersion relation of the acoustic and the first standing mode of Co layers of different thicknesses grown on Cu(001). The dispersion relation of the measured two lowest-energy magnon modes are compared to the results of the adiabatic calculations  \cite{Bergqvist2013, Costa2004}. It has been observed that both the calculations with a large interlayer coupling by Bergqvist et al., \cite{Bergqvist2013} as well as a simple nearest neighbor Heisenberg model, assuming an isotropic exchange parameter, can explain the experimental data. Note that in order to have a good agreement between theory and experiment, the calculated exchange parameters from first principles \cite{Bergqvist2013} are down-scaled to $85\%$  of their values. Interestingly, the confined magnon modes of Co films have also been probed by ISTS \cite{Balashov2014}. The measurements are performed for different thicknesses of Co films from 3 to 9 ML.  Figure \ref{Fig:DispersionCoCu} provides a comparison between the results of ISTS and the ones of adiabatic calculations \cite{Bergqvist2013}. The experimental dispersion relation shows a good agreement with the results of DFT calculations of Co films \cite{Balashov2014} and also the calculations performed for bulk Co \cite{Pajda2001}. Note that ISTS  probes the magnons with the wave vector perpendicular to the surface (the horizontal axis in Fig. \ref{Fig:DispersionCoCu} represents the wave vector perpendicular to the film surface).

\subsection{FePd alloy films on Pd(001)}

In the above discussion we have only considered ferromagnetic films with identical magnetic atoms. Now one may raise the question: How the pattern and the values of the Heisenberg exchange coupling constants change when one replace some of the magnetic atoms with nonmagnetic ones? The answer to this question would also shed light on the origin of ferromagnetism in metallic alloys, in particular to unreveal the question how the ferromagnetism evolves when one starts with a nonmagnetic lattice and dope it with ferromagnetic atoms? Investigation of magnetic excitations as a function of doping can provide detailed information regarding the evolution of HEI in alloys.
For this purpose different amount of Fe atoms are deposited on a clean Pd(001) surface. The sample is annealed at 400 K to allow the interdiffusion of Fe atoms into the topmost Pd layer. This treatment leads to the formation of a two atomic layer thick Fe$_x$Pd$_{y}$ film. If the initial coverage of Fe is one atomic layer, a two atomic layer thick Fe$_{50}$Pd$_{50}$ film will be formed \cite{Meyerheim2006}.

\begin{figure}[h!]
\begin{center}
\includegraphics[width=0.8\textwidth]{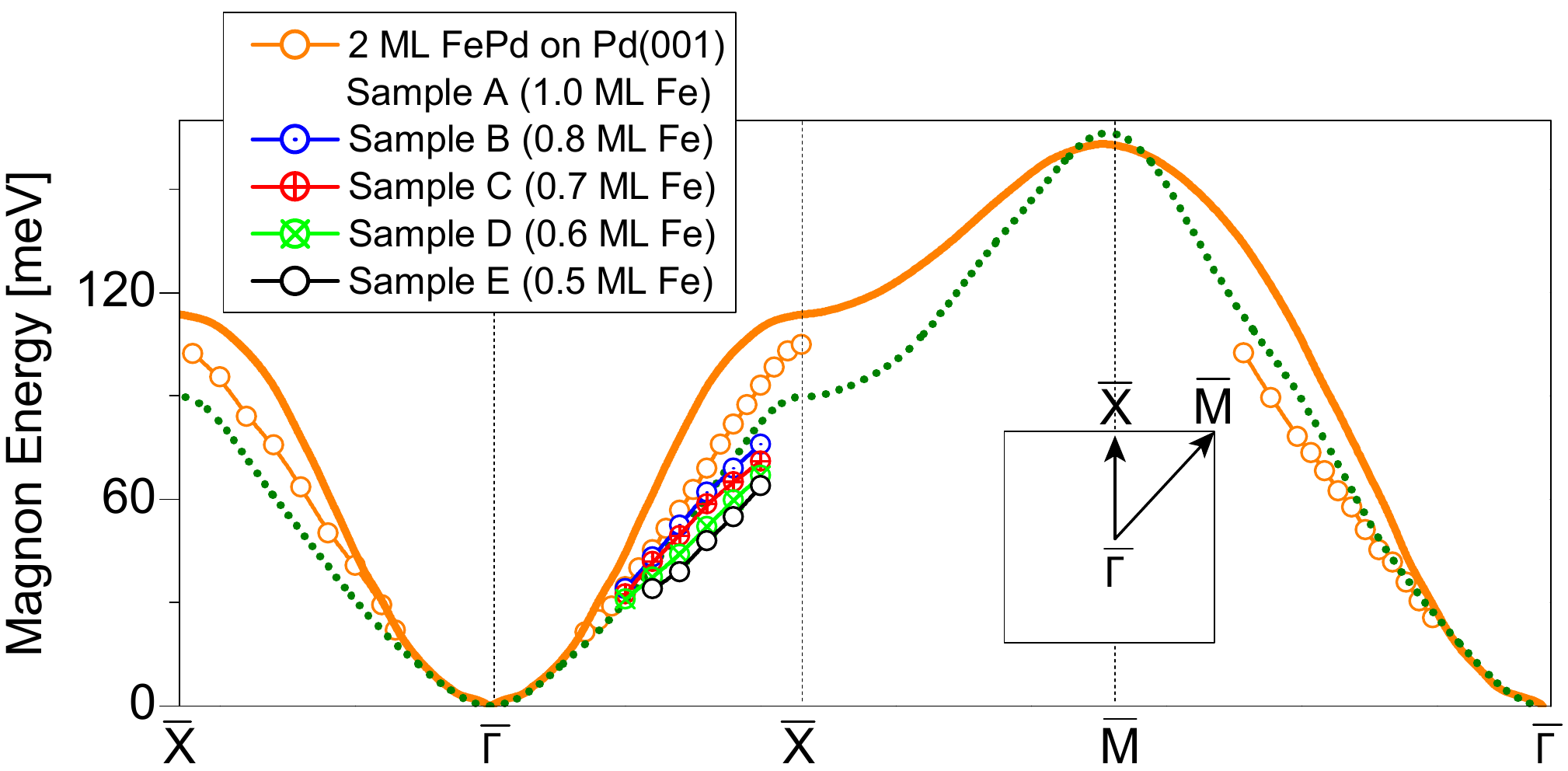}
\caption{\label{Fig:DispersionFePd}  The magnon dispersion relation of a 2 ML FePd alloy film on Pd(001). The experimental results are shown by open symbols. The results of the ab initio calculations for an Fe monolayer and a two monolayer Fe$_{50}$Pd$_{50}$ random alloy film (CPA calculations) on Pd(001) are also shown by dotted and solid curves, respectively. The corresponding exchange coupling constants are listed in Tab. \ref{Tab:ExchangeParameters}. In addition, the magnon dispersion relation measured on the samples with lower Fe amount is also presented for a comparison. In the experiment a 2 ML FePd film is prepared by depositing one monolayer Fe on the Pd(001) surface and post annealing at 400 K (sample A). Samples B, C, D, E are prepared with the same method but with lower initial coverage of Fe. Adopted from Qin et al. \cite{Qin2015}. Copyright (2015) by Macmillan Publishers Ltd: Nature Communications.}
\end{center}
\end{figure}

Figure \ref{Fig:DispersionFePd} shows the magnon dispersion relation probed by SPEELS on samples with different amount of Fe. The results of the ab initio calculations for the case of 2 ML Fe$_{50}$Pd$_{50}$ film on Pd(001) and 1 ML Fe on Pd(001) are also shown for comparison. The magnetic exchange parameters, calculated by means of the magnetic force theorem \cite{Liechtenstein1987} for both systems are listed in Tab. \ref{Tab:ExchangeParameters}. A numerical scheme based on the coherent potential approximation (CPA) \cite{Soven1967, Gyorffy1972, Theumann1974, Tang2006} has been used. It is well-known that CPA properly accounts for the alloy properties. For the case of the Fe monolayer, the HEI is confined in the plane of the monolayer. When Fe and Pd are interchanged, forming two atomic layers of FePd alloy, the interaction between the atoms sitting in different layers becomes important. Hence, both the pattern and the magnitude of the exchange coupling constants change. Ab initio calculations reveal that while the Fe-Fe intralayer interaction in the interface layer is very similar to that of the Fe monolayer on Pd(001), the Fe-Fe interlayer interaction is rather large  (see Tab. \ref{Tab:ExchangeParameters}). As a result, the magnon energies in the alloy film are higher than the ones in the Fe monolayer. Since the sample is not a perfect random alloy, some disagreements with the results of the ab initio calculations are expected.  Similar to the Fe monolayer on W(110), the exchange interaction in the Fe monolayer on Pd(001) is rather long range, meaning that the interactions beyond nearest neighbor are not negligible.

Calculations indicate that the Fe-Fe exchange parameters are considerably larger than the Fe-Pd and Pd-Pd exchange
parameters. The intralayer exchange parameters in different layers are also different. The origin of this difference lies in the stronger electronic hybridization of the interface layer with the substrate, as discussed in the previous section. This hybridization is not as strong as the one of the Fe/Ir interface. The first nearest neighbors Fe-Fe intralayer exchange parameter in the interface layer is about $14.9$ meV, very similar to the one of the Fe monolayer ($15.2$ meV). The second nearest neighbors Fe-Fe intralayer exchange parameter in this layer is about $4.3$ meV which is larger than the one of the Fe monolayer ($0.4$ meV). The third nearest neighbors intralayer exchange parameter is negative for both cases. The first nearest neighbors Fe-Fe intralayer exchange parameter in the surface layer is larger than the one in the interface layer. While the first nearest neighbors Fe-Fe interlayer exchange parameter which couples the Fe atoms from different layers is positive and rather large ($23.1$ meV), the second, third and fourth nearest neighbors interlayer exchange parameters are rather small and have a negative sign.

\begin{table}[t!]
\caption{The calculated interatomic exchange parameters for one atomic layer of Fe and two atomic layers of Fe$_{50}$Pd$_{50}$ random alloy on Pd(001). The interaction between the atoms from the same layer is referred to as intralayer interaction and the one of the atoms from different layers is referred to as interlayer interaction. The letters ``I" and ``S" denote the interface (the layer next to the substrate) and the surface layer (the layer next to the vacuum), respectively. All values are given in meV. Adopted from Qin et al. \cite{Qin2015}. Copyright (2015) by Macmillan Publishers Ltd: Nature Communications.} \label{Tab:ExchangeParameters}
\begin{center}
\begin{tabular}{c c c c c }
\hline  \hline
  System & Neighbor & Intralayer (I) & Intralayer (S) & Interlayer \\
  \hline
                                ~ & $J_{n}$ & $15.4$ & --      & --  \\
                                ~ & $J_{2n}$ & $0.4$  & --      & --  \\
1 ML Fe/Pd(001)        & $J_{3n}$ & $-1.0$ & --      & --  \\
                                ~ & $J_{4n}$ & $0.1$ & --      & --  \\
                                ~ & $J_{5n}$ & $-0.3$ & --      & --  \\
                                ~ & $J_{6n}$ & $0.8$  & --      & --  \\
  \hline
                                ~ & $J_{n}$ & $14.9$ & $24.9$ & $23.1$ \\
  2 ML Fe$_{50}$Pd$_{50}$/Pd(001) & $J_{2n}$ & $4.3$  & $1.6$  & $-0.2$ \\
                                ~ & $J_{3n}$ & $-2.1$ & $-1.9$ & $-0.6$ \\
                                ~ & $J_{4n}$ & $-0.2$ & $-0.2$ & $-0.4$ \\
  \hline  \hline
\end{tabular}
\end{center}
\end{table}

The Fe-Pd and Pd-Pd exchange parameters are much smaller than the Fe-Fe exchange parameters. For instance the first nearest neighbors Fe-Pd exchange parameter is only about $2$ meV. The main contribution to the magnon dispersion relation comes from the Fe-Fe interatomic exchange parameters. One would therefore expect lower magnon energies for samples with lower concentration of Fe, as it has been observed in the experiment (see Fig. \ref{Fig:DispersionFePd}).

\subsection{Impact of the interatomic distances on HEI in layered magnetic structures}

As the HEI describes the overlap of the electronic wavefunctions, changes in the interatomic distance have  direct consequences on the HEI.  In an atomic dimer formed by magnetic atoms, it is rather straightforward to imagine that the HEI between neighboring moments becomes weaker when increasing the distance between the magnetic atoms. However in the case of ultrathin metallic ferromagnets it is not easy to predict the change in the HEI due to the change of the interlayer distances, as the inter- and intralayer electronic wavefunctions are strongly interconnected. A change in the interlayer distance can lead to a complex reconstruction of the electronic structures. This fact has been investigated in detail for an Fe bilayer \cite{Chuang2012}. In the experiment it has been observed that magnon energies for the Fe bilayer grown on two layers of Au on W(110) are lower than the ones of Fe bilayer directly grown on W(110). This is surprising at the first glance, as the Fe-Fe interlayer distance in the latter case are larger. First principles calculations have shown that an increase of the interlayer distance of Fe layers leads to an expected decrease of the interlayer exchange parameters. This decrease of the interlayer exchange interaction is associated with the decrease of the interlayer hybridization of the electronic states. At the same time, an increase in the intralayer exchange parameters takes place. The opposite trend of the inter- and intra-layer exchange parameters is a consequence of strong redistribution of the electronic states due to the change of the interlayer distance. A careful analysis of the orbital-resolved density of states of the atoms sitting in different layers has shown that the $3d$ states responsible for the interlayer hybridization (orbitals which are extended along the film normal $z$ i.e. $d_{xz}$, $d_{yz}$, and $d_{z^2}$ orbitals) and those responsible for the intralayer hybridization (orbitals which are expanded in each layer i.e. $d_{x^2-y^2}$ and $d_{xy}$ orbitals) respond differently to the change of the interlayer distance. With increasing interlayer distance all $3d$ states (both majority and minority states) shift to lower energies. This is due to the decreased  band width of $3d$ states. In the case of  $d_{x^2-y^2}$ and $d_{xy}$ orbitals, however, some additional states appear in the spin-down (minority) density of states near the Fermi level. The appearance of these states results in an enhanced overlap of the $xy$ orbitals which in turn leads to an increase of the intralayer exchange coupling. This is a clear indication that the inter- and intralayer exchange coupling constants in layered structures composed of metallic ferromagnets are strongly interconnected \cite{Chuang2012}.

\subsection{Direct evidence of antiferromagnetic exchange interaction in ferromagnetic films} \label{Sec:DirectEvidAFM}

We discussed in Sec. \ref{Sec:FeMLW} that ab initio calculations have predicted that the intralayer exchange coupling in an Fe(110) monolayer on W(110) is of antiferromagnetic nature. However this negative exchange coupling has no significant effect on the magnon dispersion relation. This is partially due to the symmetry of the bcc(110) surface. The negative exchange coupling constant is compensated by the presence of positive higher order interactions. Therefore it is not possible to directly see the consequence of the antiferromagnetic exchange interaction on the magnon dispersion relation at the first glance. In the case of tetragonally distorted Fe(001) and FeCo(001) films it has been shown that the negative exchange coupling constants lead to an anomalous softening of the magnons at high symmetry points of the surface Brillouin zone. In Secs. \ref{Sec:Fe_Rh} and \ref{Sec:FeCo_Ir} we discuss the results of those investigations.

\subsubsection{The case of tetragonally distorted Fe films on Rh(001)}\label{Sec:Fe_Rh}

\begin{figure}[h!]
\begin{center}
\includegraphics[width=0.5\textwidth]{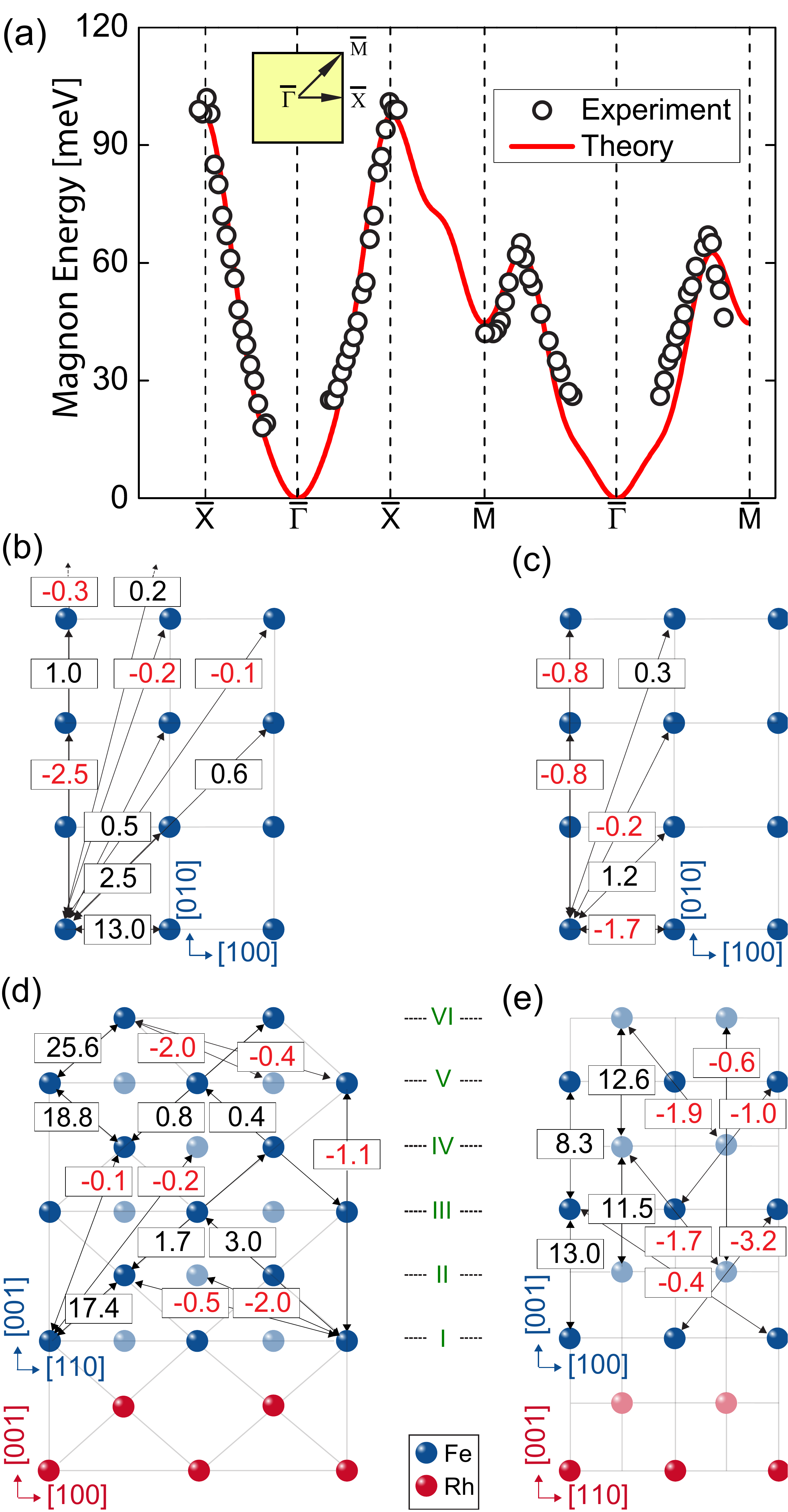}
\caption{\label{Fig:DispersionFe_Rh}  (a) The experimental acoustic magnon dispersion relation plotted together with the results of ab initio calculations. Inset shows the surface Brillouin zone. The values of the exchange parameters are given in (d)--(e). The intralayer exchange parameters in the surface layer (enumerated as VI) and in the interface layer (enumerated as I) are provided in (b) and (c), respectively. The interlayer exchange parameters are presented in (d) and (e). In (d) the Fe atoms in the Fe(110)-plane and in (e) the Fe atoms in the Fe(100)-plane with the corresponding exchange coupling constants are depicted. The light color represents the atoms in the lower atomic plane. The values are given in meV. Adopted from Meng et al. \cite{Meng2014}. Copyright (2014) by the American Physical Society.}
\end{center}
\end{figure}

The magnon dispersion relation measured on a 6 ML Fe film epitaxially grown on Rh(001) is shown in Fig. \ref{Fig:DispersionFe_Rh}. Along the $\bar{\Gamma}$--$\bar{\rm{X}}$ direction, the magnon energy increases as the wave vector increases. It reaches the maximum value of 102 meV  at the zone boundary ($\bar{\rm{X}}$-point). The magnon energy decreases while further increasing of the wave vector. This is the indication of reaching the second Brillouin zone. Such an observation is expected due to the translational symmetry of the two-dimensional Brillouin zone. The key observation is the unusual behavior of the magnon energy versus wave vector along the $\bar{\Gamma}$--$\bar{\rm{M}}$ direction. The magnon energy first increases with increasing the wave vector up to about $Q=1.15$ \AA$^{-1}$, then it decreases when further increasing the wave vector. This is unexpected as 1.15 \AA$^{-1}$ is not the zone boundary. Based on Eq. (\ref{Eq:DispersionRelation}) one can simply show that such a peculiar behavior is not expected for normal ferromagnets with ferromagnetic exchange interaction between all neighbors. If all the exchange parameters are positive, the energy of high symmetry $\bar{\rm{M}}$ point has to be larger than the other $Q$ points in the Brillouin zone.  The softening of magnons at the  $\bar{\rm{M}}$--point is a hallmark of a complex pattern of magnetic exchange interaction, including a large antiferromagnetic component, in the system.

First principles calculations have been utilized to quantify the antiferromagnetic HEI in the system \cite{Meng2014}. The calculated dispersion relation of the acoustic magnon mode is represented by the solid curve in Fig. \ref{Fig:DispersionFe_Rh} (a). The structural parameters used for the calculations have been taken from the available experimental data.
The calculated values of the exchange coupling constants are provided in Fig.  \ref{Fig:DispersionFe_Rh} (b), showing a very complex pattern of exchange coupling constants. For instance in the surface layer  (named as layer VI in Fig. \ref{Fig:DispersionFe_Rh}) the first ($+13$), second ($+2.5$), fourth ($+0.5$), fifth ($+0.6$), sixth ($+1$) and tenth ($+0.2$) nearest neighbors intralayer exchange constants are positive whereas the third ($-2.5$), seventh ($-0.2$) and ninth ($-0.3$) nearest neighbors intralayer exchange parameters are negative. In the interface layer (depicted as layer I in Fig. \ref{Fig:DispersionFe_Rh}), most of the exchange parameters are negative. For instance the first ($-1.7$), third ($-0.8$), fourth ($-0.2$) and sixth ($-0.8$) nearest neighbors intralayer exchange coupling constants are negative and only the second ($+1.2$) and seventh ($+0.3$) nearest neighbors ones are positive. This means that an antiferromagnetic exchange interaction is the dominating interaction in the interface layer.  The interlayer exchange coupling constants show also a rather complex pattern. While the first ($+17.4$), second ($+13$) and fifth ($+3$) nearest neighbors interlayer exchange constants describing the coupling of layer I to the top layers are positive, the third ($-3.2$), fourth ($-2$), sixth ($-0.1$), seventh ($-0.5$), eighth ($-0.4$) ninth ($-0.2$) and tenth ($-1.1$) nearest neighbors interlayer exchange constants (coupling this layer to the top layers) are negative. The interlayer exchange constants describing the coupling of layer VI to the layers below are ferromagnetic and only the third ($-1.9$), fourth ($-2$), seventh ($-0.4$) and tenth ($-0.6$) nearest neighbors interlayer exchange constants are negative. As it is obvious from Figs. \ref{Fig:DispersionFe_Rh} (b) -- (e) the exchange coupling constants at the interface are substantially smaller than the ones at the surface. As a result the acoustic magnon mode of the system presented in Fig. \ref{Fig:DispersionFe_Rh} (a) is localized at the interface, similar to that of the Fe/Ir(001) system discussed in Sec. \ref{Sec:Fe_Ir}. This has indeed been confirmed by the calculated layer-resolved susceptibility Bloch's spectral function \cite{Meng2014}.

\begin{figure}[h!]
\begin{center}
\includegraphics[width=0.6\textwidth]{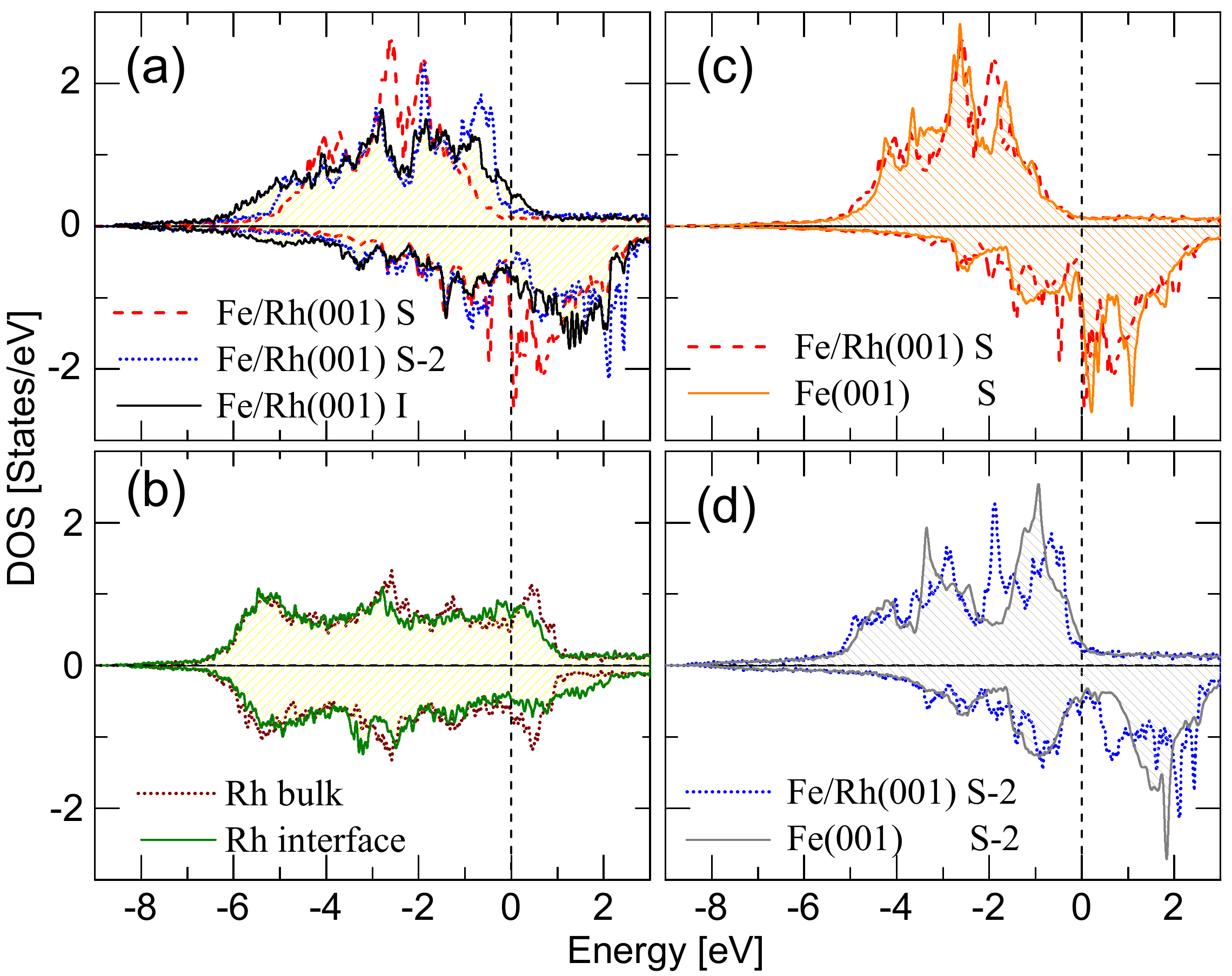}
\caption{\label{Fig:DOS} (a) Spin-resolved density of states of Fe atoms in the Fe/Rh(001) system. The states of atoms sitting in the surface layer, S (dashed curve), in the second layer below the surface, S-2 (dotted curve), and in the interface Fe layer, I (solid curve) are shown separately. (b) Spin-resolved density of states of Rh atoms sitting in the bulk Rh far from the interface (dotted curve) and in the interface Rh layer (solid curve) in the Fe/Rh(001) system.
(c) A comparison between density of states of Fe atoms sitting in the surface layer of the Fe film on Rh(001), S
(dashed curve), and in the surface layer of Fe(001), S (solid curve). (d) A comparison between density of states of Fe atoms sitting in the second layer below the surface layer of the Fe film on Rh(001), S-2 (dotted curve), and in the second layer below the Fe(001) surface, S-2 (solid curve).  Adopted from Meng et al.
\cite{Meng2014}. Copyright (2014) by the American Physical Society.}
\end{center}
\end{figure}

In order to understand the origin of the complex pattern of exchange coupling constants one should carefully analyze the
layer-, orbital- and spin-resolved electronic structures of the system. Such kinds of analysis have been performed for Fe films on Rh(001) in Ref. \cite{Meng2014} and the results are presented in Fig. \ref{Fig:DOS}. The density of states (DOS) of atoms sitting in five different places of the Fe(001)/Rh(001) system are presented and compared: ($i$) in the topmost Fe layer (surface, denoted as S), ($ii$) in the second Fe layer below the surface layer of the Fe film (denoted as S-2), ($iii$) in the interface Fe layer (denoted as I), ($iv$) in the interface Rh layer and ($v$) in the bulk Rh.
 The results are compared to those of the Fe atoms sitting at the surface and in the second layer below the surface of a semi-infinite Fe(001) with the bulk lattice constant.

 If one compares the DOS of interface Fe atoms to the DOS of atoms sitting at other places of the film, one realizes that both spin-up and spin-down states of the interface Fe atoms are spread over a larger energy range, compared to the states of the Fe atoms sitting in the other Fe layers. Moreover,  some pronounced spin-up states exist near the Fermi-level. These states do not exist in the DOS of the Fe atoms sitting in the surface layer of the Fe film on Rh(001) and also in the DOS of the Fe atoms sitting in the layer S-2 of the film (see Fig. \ref{Fig:DOS} (a)). The orbital-resolved density of states indicate that these states belong to the $\Delta_5$-symmetry orbitals. The pronounced peak in the spin-up states of the surface Fe atoms at $-1.9$  is of $d_{z^2}$ and the one at $-2.7$ eV are of $d_{xz}$ ($d_{yz}$) character. These states are largely suppressed and slightly shifted towards negative energies in the DOS of the interface Fe atoms. The spin-down DOS of the interface Fe atoms are different from the DOS of the Fe atoms sitting in the surface layer. The sharp minority surface state, located just slightly above the Fermi-level in the spin-down (minority) DOS of the Fe atoms sitting in the surface layer, is absent in the case of the Fe atoms sitting in the interface Fe layer. Moreover, the peak at about 0.7 eV, is shifted to 1.4 eV and is broadened. This peak is mainly of $d_{z^2}$ character.

Now if one compares the DOS of the interface Rh atoms with the ones of the bulk Rh, one recognizes that the spin-up states in the interface Rh atoms are at lower energies compared to the ones of the Rh atoms sitting in bulk Rh (Fig. \ref{Fig:DOS} (b)). The spin-down states of the interface Rh atoms are at higher energies with respect to the states of the bulk Rh atoms. A large upward shift of about $1.5$ eV is occurred for the spin-down states which are of $d_{z^2}$ character. The sharp states at about $-3.2$ eV in the DOS of the interface Rh atoms are caused by a downward shift of the states located at about $-2.6$ eV in the DOS of the bulk Rh atoms. These states are of $d_{x^2-y^2}$ and $d_{xz}$ ($d_{yz}$) character.
The considerably large states at about $-2.5$ eV in the spin-down states of the interface Rh atoms are rather small in the spin-down DOS of the bulk Rh atoms. These sates are of $d_{xy}$, $d_{xz}$ and $d_{yz}$ character.
All the differences in the DOS of the interface Rh atoms with respect to the DOS of Rh atoms in the bulk Rh indicate a strong hybridization of the electronic states at the interface, which alters to a large extent the HEI across the film and in particular at the interface.

Now one may compare the DOS of Fe atoms located in the surface layer of the Fe films on Rh(001) to the DOS of Fe atoms located at the surface of an Fe(001) semi-infinite slab with bulk lattice parameter (Fig. \ref{Fig:DOS} (c)). Such a comparison would reveal the importance of the tetragonal distortion on the negative exchange coupling constants. As seen in Fig. \ref{Fig:DOS} (c), the large spin-up states at the energy of about $-1.65$ eV in the DOS of the Fe atoms sitting in the surface layer of Fe(001) are shifted to $-1.9$ in the DOS of the Fe atoms in the surface layer of the Fe film on Rh(001). These sates are of $d_{z^2}$ character. The spin-up states at $+1$ eV in the DOS of the Fe atoms at the Fe(001) surface are of the same character but are shifted to $+0.7$ eV in the DOS of the Fe atoms sitting at the surface of the Fe film on Rh(001). The minority surface state of the Fe(001) surface which is usually located just above the Fermi-level is slightly shifted downwards in the case of the tetragonally distorted Fe(001) film on Rh(001). The differences in both the spin-up and spin-down density of states of the atoms sitting in the layer S-2 (second layer below the surface) compared to the states of the atoms sitting in the same layer of the semi-infinite Fe(001) are caused by a shift of about $0.3$ eV of the $d_{x^2-y^2}$ orbitals and a shift of about $0.6$ eV of the $d_{xz}$ and $d_{yz}$ orbitals.

\begin{figure}[h!]
\begin{center}
\includegraphics[width=0.99\textwidth]{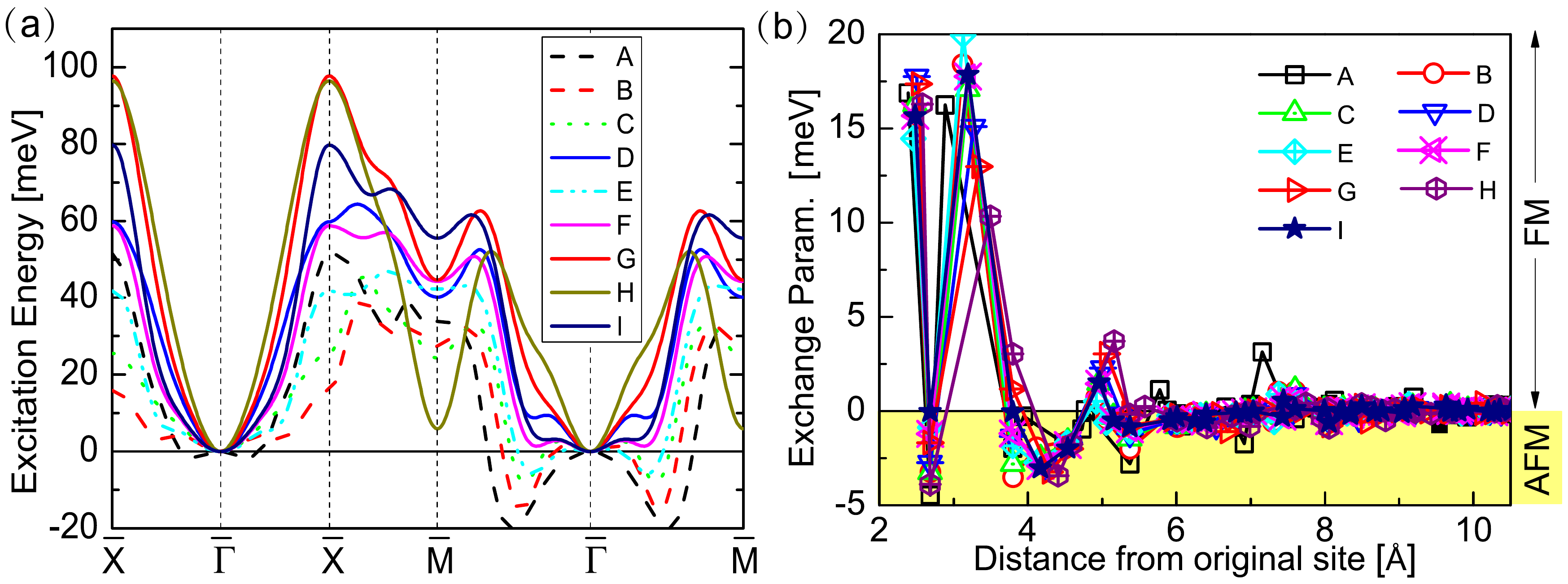}
\caption{\label{Fig:DispersionFeRd_differentCases} (a) The magnon dispersion relation calculated for 6 ML Fe on Rh(001) using different values of interlayer distances as given in Tab. \ref{Tab:Cases}. For all cases a magnon softening at the high symmetry $\bar{\rm{M}}$--point has been observed. (b) The calculated Fe--Fe exchange parameters where the first Fe atom is sitting in the interface layer and the second one is sitting in the distance $x$ from the first one.  For all cases a considerably large antiferromagnetic exchange interaction has been observed.  Adopted from Meng et al. \cite{Meng2014}. Copyright (2014) by the American Physical Society.}
\end{center}
\end{figure}

\begin{table}[t!]
\caption{The calculated values of the magnon softening at the $\bar{\rm{M}}$-point for different values of interlayer spacing for the Fe/Rh(001) system. The values of the Fe-Rh and Fe-Fe interlayer spacing are given in column 2 and 3, respectively. Column 4  shows the ratio of the height of the unit cell divided by in-plane lattice parameter ($c/a=1$ corresponds to a cubic structure). The last column represents the ratio of the magnon energies at $\bar{\rm{M}}$- and $\bar{\rm{X}}$-points, indicating the $\bar{\rm{M}}$-point softening. \label{Tab:Cases}}
\begin{center}
\begin{tabular}{c c c c c }
\hline  \hline
  Case & $d_{Fe-Rh}$  [\AA] & $d_{Fe-Fe}$  [\AA]& $c/a$ & $\varepsilon_{\bar{M}}/\varepsilon_{\bar{X}}$ \\
  \hline
  A & $1.44$ & $1.44$ & 1.07      & 35/50  \\
  B & $1.56$ & $1.56$  & 1.16      & 30/15  \\
  C & $1.60$ & $1.60$ & 1.19      & 25/25  \\
  D & $1.63$ & $1.63$ & 1.22     & 40/60  \\
  E & $1.71$ & $1.56$ & 1.16      & 42/41  \\
  F & $1.71$ & $1.60$  & 1.19      & 45/60  \\
  G & $1.74$ & $1.66$ & $1.24 $ & 45/98 \\
  H & $1.74$ & $1.74$  & $1.30$  & 6/98 \\
   I & $1.74$ & $1.60$ & $1.30$ & 65/60 \\
    \hline  \hline
\end{tabular}
\end{center}
\end{table}

The observed peculiar magnon softening at the $\bar{\rm{M}}$--point is due to both the tetragonal distortion of the film and also the presence of the Rh substrate. The calculations performed for different values of the  Fe-Fe and Fe-Rh layer spacing could address the  influence of the film structure as well as the electronic hybridization with the substrate on the magnon dispersion relation and the softening at the $\bar{\rm{M}}$--point. Calculations have been performed for different cases listed in Tab. \ref{Tab:Cases} and the results are presented in Fig. \ref{Fig:DispersionFeRd_differentCases} (a).
The calculated values of the exchange coupling constants are plotted in Fig. \ref{Fig:DispersionFeRd_differentCases} (b.)
The magnon softening at the $\bar{\rm{M}}$--point and the negative exchange coupling constants have been observed for all cases.  For unrealistic values of interlayer spacing (cases A, B,C and E) negative values of magnon energies
in the midway of $\bar{\Gamma}$-$\bar{\rm{M}}$ direction have been observed. For the cases F, H and I the antiferromagnetic HEI shows up in the magnon dispersion relation as softening at the $\bar{\rm{M}}$ point. The
experimental magnon dispersion relation fits the case G in which the experimental values of the interlayer spacing are taken into account as the input of the ab initio calculations. The calculations have further shown that changing the value of the interlayer distance between the first Fe layer and Rh ($d_{Fe-Rh}$) by $\pm$0.04~\AA~does not change the overall
shape of the magnon dispersion relation. It only slightly changes the absolute value of magnon energies. The main experimental observation i.e. the softening at the $\bar{\rm{M}}$--point remains unaffected.
This means that even without the first principles calculations and solely based on the observed magnon softening at the $\bar{\rm{M}}$--point one can conclude the existence of a large antiferromagnetic exchange interaction in
the system. The observed unusual magnon softening at the $\bar{\rm{M}}$ point can be regarded as a signature indicating the presence of an antiferromagnetic exchange coupling in the system.

Based on static magnetic measurements different magnetic structures are suggested for the Fe and FeCo films on Rh(001). For example Hwang et al. have suggested an AFM \cite{Hwang1999} ground state for ultrathin Fe films below 4 ML while Hayashi et al. have suggested the presence of a magnetic ``dead layer" \cite{Hayashi2001, Hayashi2004} at the interface. Scanning tunneling microscopy and spectroscopy investigations by Takada et al. have shown that the magnetic unit cell of Fe monolayer on Rh(001) is rather complex, indicating a complex noncollinear AFM ground state \cite{Takada2013}. Similarly a complex  spin structure is suggested for FeCo/Rh multilayers \cite{Yildiz2009, Przybylski2012} on Rh(001) based on the analysis of the magneto-optical Kerr effect and polarized soft x-ray resonant magnetic reflectivity data. Theoretical calculations have predicted that the AFM configuration is favored for the Fe monolayer \cite{Spisak2006, Al-Zubi2011}. The direct measurements of the exchange coupling parameters via magnon spectroscopy indicates that a large antiferromagnetic exchange interaction is present in the system, which can lead to a complex AFM ground state depending on the film thickness.

The important message here is that the pattern of the magnetic exchange parameters in a $3d$ ferromagnetic film on a nonmagnetic substrate can be extremely complicated. Although the film may show a typical ferromagnetic hysteresis loop, the HEI across the film can be of both FM and AFM nature. Such a complicated pattern of the exchange coupling constants can be quantified by looking at the dynamics of the system i.e. by magnon spectroscopy.

\subsubsection{The case of tetragonally distorted Fe/Co multilayers on Rh(001) and Ir(001)} \label{Sec:FeCo_Ir}

\begin{figure}
\begin{center}
\includegraphics[width=0.45\textwidth]{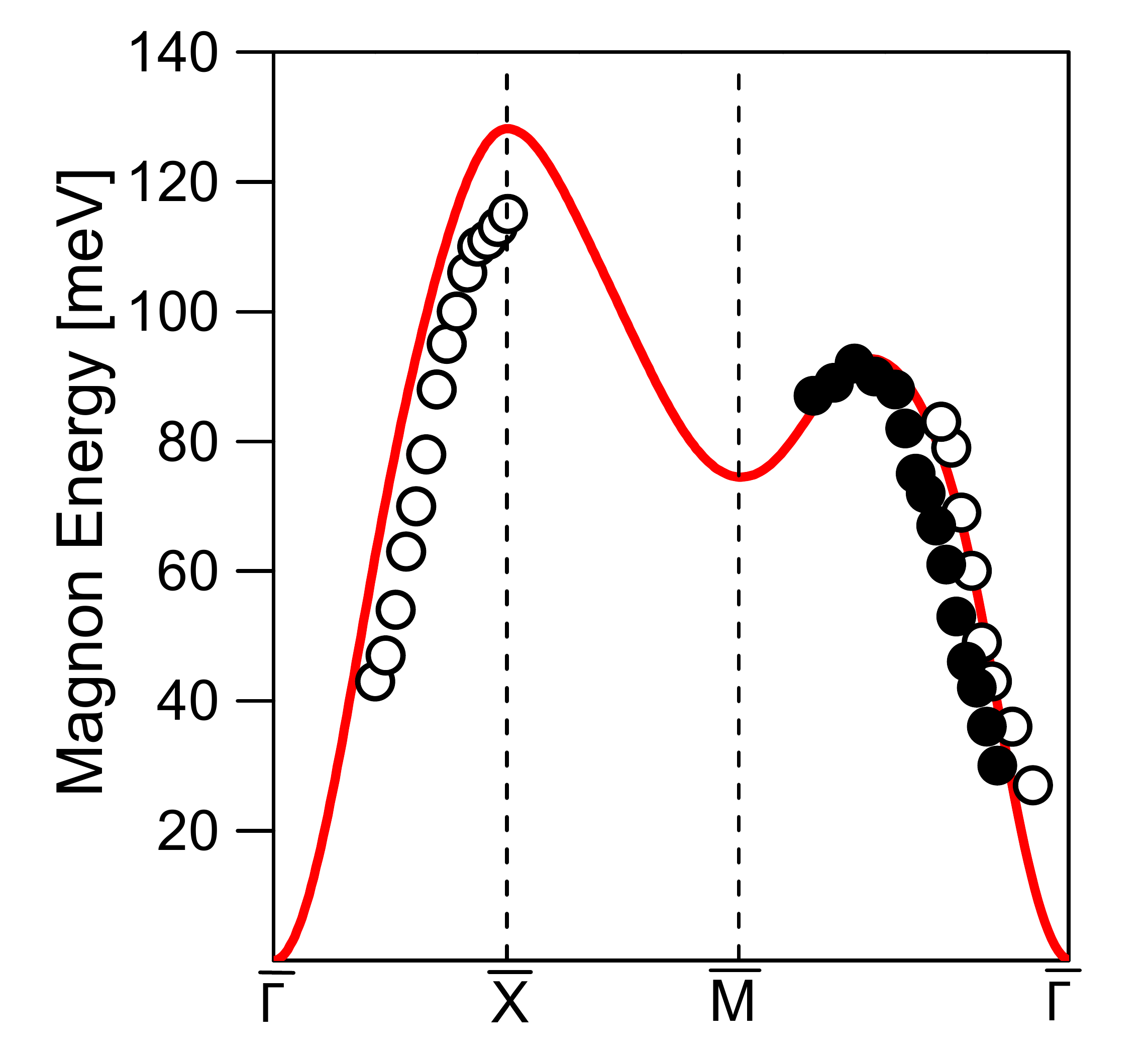}
\caption{\label{Fig:DispersionCoFe_Ir_2}  The acoustic magnon dispersion relation of 2 ML Co on 1 ML Fe on Ir(001). The experimental results are shown by open and solid circles and the results of the ab initio calculations are presented by the solid curve. The data shown by open symbols are the ones shown in Fig. \ref{Fig:DispersionCoFe_Ir}. The new data, shown by the solid circles, are measured up to the $\bar{\rm{M}}$--point. The magnon softening at this high symmetry point is a result of the antiferromagnetic intralayer exchange interaction in the Fe monolayer buried under the Co layers. The corresponding values of the exchange parameters are listed in Fig. \ref{Fig:DispersionCoFe_Ir} (c).}
\end{center}
\end{figure}

Ab initio calculations predict a magnon softening in the case of FeCo multilayers on both Rh (001) and Ir(001) surfaces (see the solid curve in Fig. \ref{Fig:DispersionCoFe_Ir} (a)). For the case of FeCo multilayers on the Ir(001), in the earlier experiments the data points close to the $\bar{\rm{M}}$--point were missing. The system has therefore been revisited experimentally.  The new experimental data along the $\bar{\Gamma}$--$\bar{\rm{M}}$ up to the $\bar{\rm{M}}$--point are shown in Fig. \ref{Fig:DispersionCoFe_Ir_2}. A clear magnon softening at the $\bar{\rm{M}}$--point is observed which is mainly due to the antiferromagnetic nearest neighbors intralayer coupling (see Fig. \ref{Fig:DispersionCoFe_Ir} (c)).

 Ab initio calculations of the magnon dispersion relation of tetragonally distorted bulk FeCo compound considering three different $c/a$ ratios (1.13, 1.18, and 1.24) have shown also a magnon softening at the $\rm{M}$--point  as a result of the antiferromagnetic exchange coupling constants. The results of those calculations are summarized in Tab. \ref{Tab:FeCo}.  The larger the tetragonal distortion the softer the magnons at the $\rm{M}$--point (see the last column of Tab. \ref{Tab:FeCo}). The physical origin of this antiferromagnetic exchange coupling is the same as the one of the tetragonally distorted Fe films on Rh(001), as discussed in the previous section.

\begin{table}[t!]
\caption{The magnon softening at the $\rm{M}$--point observed in the calculation of the magnon dispersion relation of tetragonally distorted bulk FeCo material. The calculations have been performed for bulk FeCo. The lattice parameters used for the calculations are the corresponding lattice parameters of FeCo films on the Pd(001), Ir(001) and Rh(001) substrates. The data are taken from Ref. \cite{Sasioglu2013}} \label{Tab:FeCo}
\begin{center}
\begin{tabular}{c c c}
\hline  \hline
  Case &  $c/a$ & $\varepsilon_{M}/\varepsilon_{X}$ \\
  \hline
  FeCo bulk alloy on Pd  &1.13 &155/210  \\
  FeCo bulk alloy on Ir    &1.18 & 80/130  \\
  FeCo bulk alloy on Rh &1.24 & 25/105  \\
\hline  \hline
\end{tabular}
\end{center}
\end{table}

\section{Direct probing of the interfacial Dzyaloshinskii--Moriya exchange interaction} \label{Sec:ProbingDMI}

The magnetic energy contribution associated with DMI in bulk $3d$ ferromagnets is zero. This is due to the following reasons. Firstly, the spin--orbit interaction in bulk $3d$ ferromagnets is very small. Secondly, the bulk $3d$ ferromagnets possess centro-symmetric structures. However, in the case of ultrathin ferromagnetic films of $3d$ ferromagnets grown on heavy-element substrates the spin--orbit coupling is no longer a small perturbation. The inversion symmetry is also broken at the surface/interface. As a matter of fact it has been demonstrated that a strong spin--orbit coupling in the presence of the broken symmetry leads to DMI, which stabilizes a noncollinear spin structure for a Mn monolayer grown on the W(110) \cite{Bode2007} and W(100) \cite{Ferriani2008} surfaces. It has further been shown that DMI is also active in the Fe monolayer and bilayer on W(110) and plays a crucial role in the determination of the magnetic ground states of these systems \cite{Vedmedenko2007,Heide2008,Heide2009, Meckler2009}. In this section we discuss how the dynamical excitations are influenced by DMI and how the magnon spectroscopy maybe used to quantify the strength of this interaction.

\subsection{Fe monolayer on W(110)}

As discussed in Sec. \ref{Sec:DMI} the presence of DMI in the system shall break the degeneracy of the magnons with opposite wave vectors. A model system for investigating the impact of DMI on the magnon dispersion relation is the Fe monolayer on W(110).  Based on ab initio calculations Udvardi and Szunyogh have predicted that the influence of the DMI on the magnon dispersion relation of an Fe monolayer on W(110) is strong enough to be observed experimentally \cite{Udvardi2009}.  Another candidate for such an investigation is the Fe bilayer on the same surface. The physics is very much the same as the one of the Fe monolayer. In the next section we discuss in detail the first experimental evidence of the influence of DMI on the magnon dispersion in an Fe bilayer on W(110). We discuss how the components of the DM vector can be quantified by probing the DMI induced energy asymmetry of magnons.

\subsection{Fe bilayer on W(110)}

In the SPEELS experiments the spectra are usually recorded at a fixed wave vector for two possible spin orientations of the incoming electron beam \cite{Plihal1999,Vollmer2003,Tang2007,Prokop2009,Zakeri2014}. The conservation of the angular momentum during the scattering prohibits magnon excitations for incoming electrons with a spin polarization antiparallel to the sample magnetization. Only electrons having spin parallel to the sample magnetization  can create magnons. The electrons with spin antiparallel to the magnetization can annihilate the thermally excited magnons and gain energy (see Sec. \ref{Sec:SPEELS} for more detail). These facts lead to a peak in the minority spin spectra ($I_{\downarrow}$) in the energy loss region and a peak in the majority spin spectra  ($I_{\uparrow}$) in the energy gain region, in analogy to the Stokes and ani-Stokes peaks in the BLS experiments. At room temperature the amplitude of the peak due to the magnon annihilation (in the energy gain region) is much smaller than the one caused by the magnon creation. This gives rise to a pronounced peak in the energy loss region and a small dip in the energy gain region of the difference spectra defined as $I_{\rm{Diff.}}=I_{\downarrow}-I_{\uparrow}$.  The dip in the gain region is small. This is due to the fact that magnons are bosons and obey the Bose-Einstein statistics. The ratio of the loss to gain amplitude is given by the Boltzmann factor.  Both gain and loss features can be better seen when plotting the asymmetry spectra $Asy.=\frac{I_{\downarrow}-I_{\uparrow}}{I_{\downarrow}+I_{\uparrow}}$ .  In the absence of DMI the magnon dispersion has to be symmetric. This means that the position of the peak and dip in the loss and gain region has to be at the same energy. In addition if one measures the magnon spectra for negative and positive wave vectors, the spectra have to be identical.

\begin{figure}
\begin{center}
\includegraphics[width=0.45\textwidth]{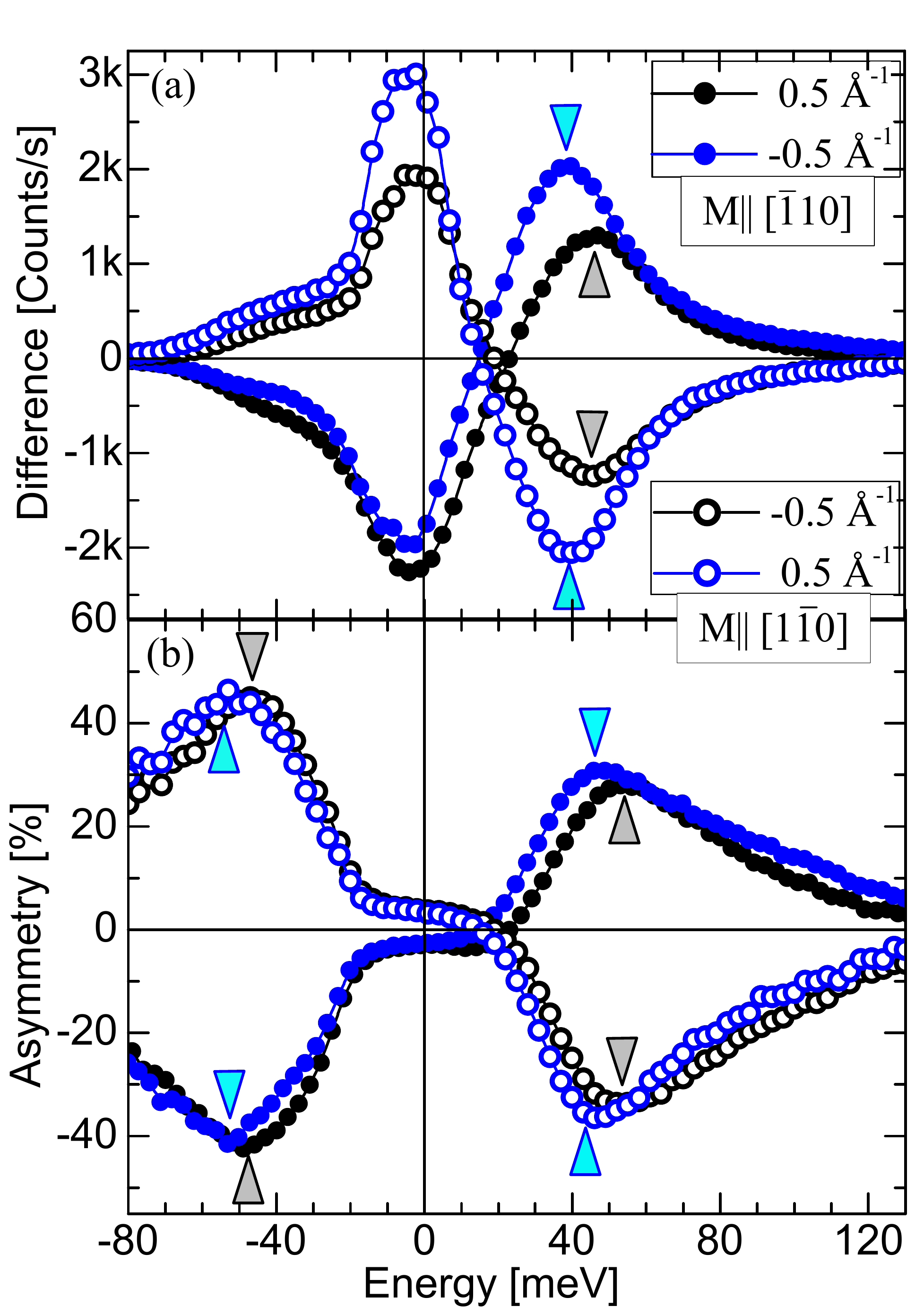}
    \caption{\label{Fig:SpectraBI} (a) Difference, $I_{\rm{Diff.}}=I_{\downarrow}-I_{\uparrow}$, and (b) asymmetry, $Asy.=\frac{I_{\downarrow}-I_{\uparrow}}{I_{\downarrow}+I_{\uparrow}}$, SPEELS spectra measured at $Q= \pm 0.5 $\AA$^{-1}$, recorded on an Fe bilayer. The filled symbols are for $\mathbf{M}\parallel[\bar{1}10]$ and the open ones are for $\mathbf{M}\parallel [1\bar{1}0]$. The big triangles show the peak positions of magnon creations and annihilations, taking place at energy loss and gain, respectively. Adopted from Zakeri et al. \cite{Zakeri2010}. Copyright (2010) by the American Physical Society.}
\end{center}
\end{figure}

In Fig. \ref{Fig:SpectraBI}  the difference and asymmetry spectra measured on an Fe bilayer on W(110) are presented. The spectra have been recorded for two opposite directions of the magnetization $\mathbf{M}$. The solid (open) symbols are the results of measurements when the magnetization is pointing along the [$\bar{1}$10]-direction ([1$\bar{1}$0]-direction). Figure \ref{Fig:SpectraBI} demonstrates that for the case of positive $Q$ the magnon creation peak in the energy loss region is at higher energies, whereas the magnon annihilation peak in the gain region is at lower ones. A similar behavior can be observed when looking at the asymmetry spectra. In the case of negative $Q$ a fully opposite trend is observed meaning that the magnon annihilation peak is at higher energies whereas the magnon creation peak is at lower energies. This observation by itself can be regarded as a strong evidence of DMI. Another strong evidence comes from the measurements of the same spectra but with opposite $\mathbf{M}$. In fact reversing $\mathbf{M}$ is somewhat equivalent to a time inversion experiment. A time-reversed experiment means that the position of the source and the detector is interchanged in the scattering experiment and the electrons travel the revered path when they are scattered from the surface. The propagation direction of magnons is also reversed. In the case of reversed $\mathbf{M}$ the magnon excitation peak for negative $Q$ is at higher energies with respect to the one for positive $Q$. A similar effect happens for the gain features.  The conclusion of those results is that the presence of DMI leads to an asymmetric magnon dispersion relation when the magnons with the wave vector perpendicular to the magnetization are measured. It is worth mentioning that DMI influences not only the magnon dispersion relation, it also alters their lifetime. It has been observed that magnons with $\pm Q$ possess a different lifetime when propagating along opposite directions, perpendicular to $\mathbf{M}$ \cite{Zakeri2012a}, in line with the results of dynamical susceptibility calculations \cite{Costa2010}. This is clearly visible by looking at the linewidth broadening of the difference spectra presented in Fig. \ref{Fig:SpectraBI}. A more detailed explanation of this so called chiral damping does not belong to the main topic of the present review and may be find elsewhere \cite{Zakeri2012a}. Recently it has been discussed that both the asymmetric magnon dispersion relation and the chiral damping display striking similarities with the Rashba type of effects on fermionic quasi-particles (electrons and holes) in solids \cite{Manchon2015}.

\begin{figure}
\begin{center}
\includegraphics[width=0.45\textwidth]{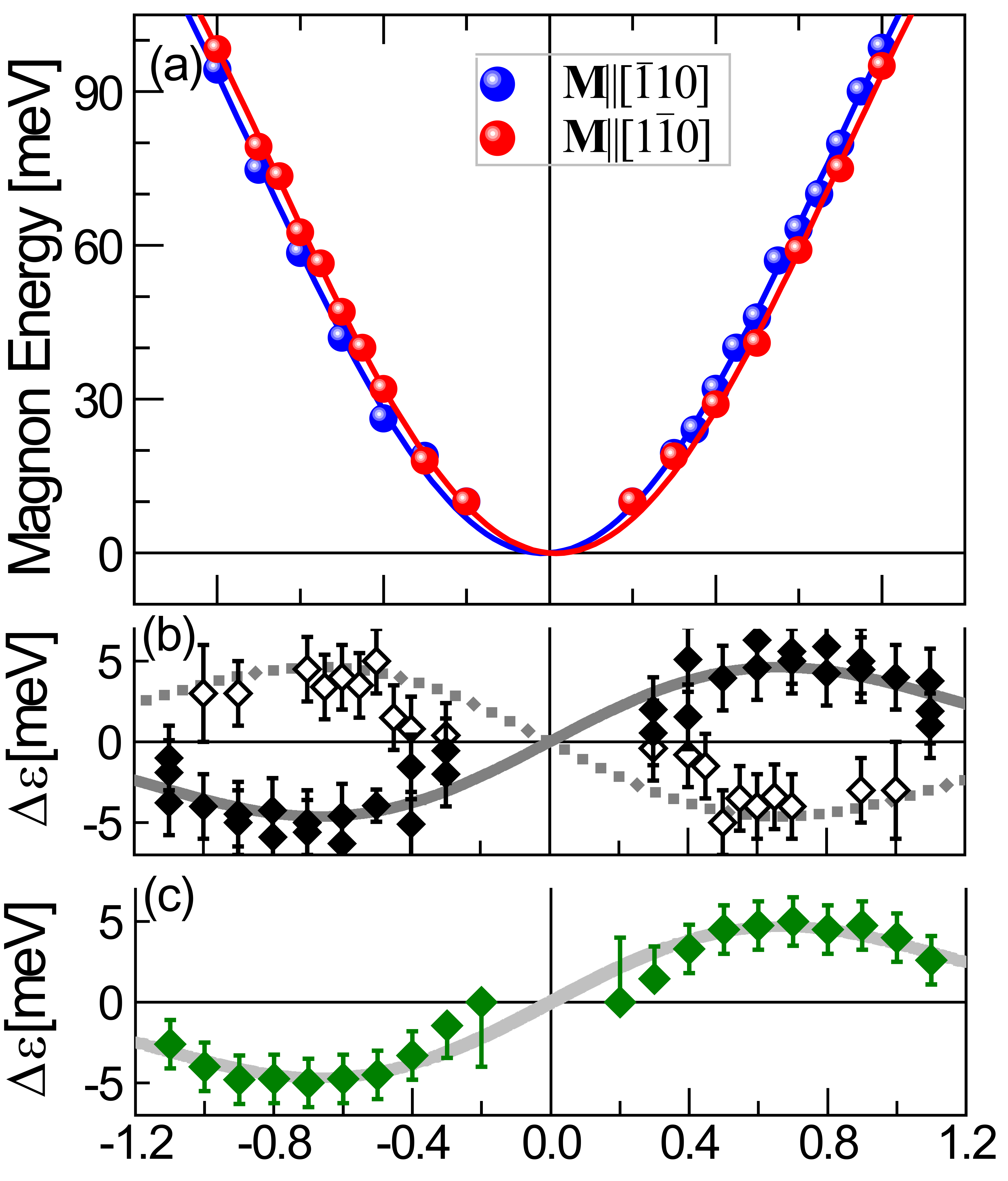}
    \caption{\label{Fig:EnergyAssy}  (a) The magnon dispersion relation measured on an Fe bilayer on W(110). The dispersion relation is measured at room temperature and for two opposite directions of magnetization being [$\bar{1}$10] and [1$\bar{1}$0] directions. (b) The energy asymmetry defined as $\Delta \varepsilon = \varepsilon(Q) - \varepsilon(-Q)$ obtained from the data presented in (a). The solid diamonds represent the experimental results of $\Delta \varepsilon = \varepsilon_{\mathbf{M}\parallel[\bar{1}10]}(Q) - \varepsilon_{\mathbf{M}\parallel[\bar{1}10]}(-Q)$, while the open diamonds represent the experimental results of $\Delta \varepsilon = \varepsilon_{\mathbf{M}\parallel[1\bar{1}0]}(Q) - \varepsilon_{\mathbf{M}\parallel[1\bar{1}0]}(-Q)$ . The solid and dotted curve represent the fit based on Eq. (\ref{Eq:DMBI}) considering different directions of $\mathbf{M}$. (c) The energy asymmetry defined as $\Delta \varepsilon = \varepsilon_{\mathbf{M}\parallel[\bar{1}10]}(Q) - \varepsilon_{\mathbf{M}\parallel[1\bar{1}0]}(Q)$ . The experimental results are shown by solid diamonds and the fit according to Eq. (\ref{Eq:DMBI}) is represented by the solid curve.  Adopted from Zakeri et al. \cite{Zakeri2010} and \cite{Zakeri2012a}. Copyright (2010) and (2012) by the American Physical Society.}
\end{center}
\end{figure}

While investigating a thin ferromagnetic film, one can, in principle, measure the magnon dispersion relation for all the possible scenarios. The possible scenarios are (i) keeping the direction of $\mathbf{M}$ fixed and probing the magnons with positive and negative $Q$ and (ii) reversing the direction of $\mathbf{M}$ and probing again the magnons with $\pm Q$. A summery of such measurements is provided in Fig. \ref{Fig:EnergyAssy}. In order to quantify DMI the energy asymmetry should be obtained from the experimental data. Starting with Eq. \ref{Eq:DispersionRelationDM} and considering the symmetry of the Fe(110) surface unit cell one can derive the following equation for $\Delta \varepsilon$:

\begin{eqnarray}
\Delta \varepsilon=\pm4c\left[\left(2D^x_{1}+\acute{D^x_{1}}\right) \sin\left(\frac{Qa}{2}\right) + D^x_{2} \sin \left(Qa\right) \right]. \label{Eq:DMBI}
\end{eqnarray}

Here the $\pm$ sign represent the different directions of $\mathbf{M}$, $a$ is the in-plane lattice constant of Fe which is the same as the one of the W substrate (3.16 \AA)~and $D^x_i=\sin^2 \theta \mathbf{D_i}\cdot \mathbf{\hat{e}}$ ($\acute{D}^x_{i}=\sin^2 \theta \mathbf{\acute{D}_i}\cdot \mathbf{\hat{e}}$) is the longitudinal component of the DMI vector of the $i^{th}$ neighbors in the same atomic plane (in the neighboring atomic plane).  Equation (\ref{Eq:DMBI}) implies that $\Delta \varepsilon=\varepsilon_{\mathbf{M}}(Q)-\varepsilon_{\mathbf{M}}(-Q) = \varepsilon_{\mathbf{M}}(Q)-\varepsilon_{\mathbf{-M}}(Q) $. This means that in practice the DMI induced energy asymmetry can be measured either by keeping the direction of $\mathbf{M}$ fixed and comparing the magnon energy of $Q$ with the one of $-Q$ or performing the measurements for two opposite magnetization directions and compare the results. Moreover, the experimental data shall show an extremum  at $\pm0.5<|Q|<\pm1$ \AA. The experimental data of energy asymmetry are presented in Figs. \ref{Fig:EnergyAssy} (b) and (c). The data are fitted by Eq. (\ref{Eq:DMBI}) and the values of  $|2D^x_1+\acute{D^x_{1}}|$=0.9(3) meV and $|D^x_2|$=0.5(3) meV have been obtained. Interestingly, the experimental data could only be fitted by assuming  $c=+1$. This means that the non-collinear spin structured favored by DMI shall have a right rotating sense. The fact that the spin spiral ground state of the Fe bilayer on W(110) is a right rotating spiral has been experimentally confirmed by means of spin-polarized scanning tunneling microscopy experiments \cite{Meckler2009}.
The DMI vectors can be calculated from first principles (for details on the methodology see for example \cite{Udvardi2003,Udvardi2009,Vedmedenko2007,Katsnelson2010,Bergqvist2013}). A comparison between the calculated values of the components of the DMI vector for Fe monolayer and bilayer on W(110) and the values reported experimentally is provided in Tab. \ref{Tab:DMI}.

\begin{table}[t!]
\caption{Components of the DMI vector in the Fe/W(110) system. The experimental results are compared to the results of first principles calculations. The values are given in meV.} \label{Tab:DMI}
\begin{center}
\begin{tabular}{l c c c}
\hline  \hline
 ~& bilayer (Exp.) \cite{Zakeri2010, Zakeri2012a} &  bilayer (Cal.) \cite{Bergqvist2013, Etz2015} & monolayer (Cal.) \cite{Udvardi2009} \\
  \hline
$|D^x_1|$ (Int.) & -- & 0.52 & 1.42\\
$|D^x_2|$ (Int.) & $ 0.5\pm0.3$ & 0 & 6.08 \\
$|\acute{D^x_{1}}|$  & -- & 0 & --\\
$|\acute{D^x_{2}}|$  & -- & 0.35 & --\\
$|D^x_1|$ (Suf.) & -- & 0.16 & --\\
$|D^x_2|$ (Surf.) & $ 0.5\pm0.3$ & 0 & -- \\
$|2D^x_1+\acute{D^x_{1}}|$  & $0.9 \pm$ 0.3 & 0.68 & -- \\

  \hline  \hline
\end{tabular}
\end{center}
\end{table}

Equation (\ref{Eq:DMBI}) may be approximated for small values of $Q$ to

\begin{eqnarray}
\Delta \varepsilon= \pm D_{DMI} Q ;~~~~~~~~~ |D_{DMI}|=4\left[\left(\frac{2D^x_{1}+\acute{D^x_{1}}}{2}\right) + D^x_{2} \right] a. \label{Eq:DMBI_2}
\end{eqnarray}

Using the values of $D^x_{i}$, discussed above and assuming $a=3.16$ \AA, one obtains $D_{DMI}=12.0 \pm1.8$ meV\AA $=2.9 \pm 0.5$ THz\AA (1 meV=0.24 THz). A similar value can be obtained when fitting the experimental data in Figs. \ref{Fig:EnergyAssy} (b) and (c) up to $Q=0.6$ \AA$^{-1}$ with a line. A comparison among the experimental $D_{DMI}$ and the results obtained by ab initio calculations is provided in Tab. \ref{Tab:DMI}.

The method described above is the most direct and straightforward way of probing the DMI vector, as it relies only on the direct measurement of the DM energy. The great advantages of this method with respect to the static measurements are, firstly, both the symmetric HEI and DMI can be quantitatively determined in one experiment. Secondly, as these two interactions influence the magnon dispersion relation differently, one can quantify one without knowing the other. This can be done without performing any complex calculation and data analysis. The DMI term is simply proportional to the energy asymmetry measured in the experiment. In the static measurements based on the analysis of domain walls one first needs to describe all the energy terms, including the DMI energy, precisely. The magnetic ground state can then be obtained by minimizing the magnetic free energy of the system. Only a correct description of the magnetic ground state leads to the quantification of DMI.

As discussed above DMI leads to the fact that the transversal magnons (the magnons with the propagation direction orthogonal to the magnetization) possess different lifetime and amplitude when propagating along opposite directions. These differences in the lifetime and amplitude together with differences in the group velocity $\mathbf{v_g} = \frac{1}{\hbar} \left(\partial \varepsilon/ \partial Q \right) \mathbf{\hat{Q}}$, caused by the asymmetric magnon dispersion relation, lead to a substantial difference in the propagation behavior of magnons with the same energy along two opposite directions. The asymmetric propagation of magnons' wave packet can be considered as another evidence of the presence of DMI. This effect opens a possibility to probe DMI by probing the magnon wave packets by means of experimental methods which enable probing the magnons in real time and space, as suggested in Ref. \cite{Zakeri2012a}.

Recently, low wave-vector magnon excitations by means of microwave \cite{Lee2015, Nawaoka2015} or light scattering \cite{Nembach2015, Cho2015, Di2015} have been utilized to address the strength of DMI. In the BLS experiments, the Stokes and anti-Stokes modes are found to appear at different frequencies. The measured frequency asymmetry of the magnons with $Q$ and $-Q$  ($\Delta \omega = \omega (Q) - \omega (-Q)$) is found to scale linearly with the wave vector. This frequency shift has been used to quantify the strength of DMI. There are a few important points which one has consider, when investigating DMI by using low energy excitations. First, in thin ferromagnetic films the presence of the dipolar interaction and the magnetic anisotropy can lead to the unidirectional Damon-Eshbach surface mode \cite{Damon1961,Grunberg1985}. Since DMI is present in the systems with broken inversion symmetry,  the interface magnetic anisotropy at the top and bottom interfaces is different. This difference in the interface magnetic anisotropy can, in principle, cause an energy (frequency) asymmetry $\Delta \omega$. Hence, a precise determination of DMI requires a careful consideration of the nonreciprocal behavior of Damon-Eshbach mode and the influence of the interfacial magnetic anisotropy. Second, the low wave vector excitations are dominated by the dipolar interactions and magnetic anisotropy. As we discussed in Sec. \ref{Sec:DMI} the quantum mechanical origin of DMI is somewhat similar to the symmetric HEI and hence one needs to investigate the large wave vector excitations (the excitations with short wave length) in order to be able to provide a true microscopic picture of DMI. The linear $Q$ dependence of $\Delta \omega$ is no longer valid for large Q values, as can be seen from the data presented in Fig. \ref{Fig:EnergyAssy}. Third, DMI is an antisymmetric interaction which should be best represented by a matrix (or at least with a vectorial) quantity. Such a physical quantity can only be well described when all the components are precisely measured. In the interpretation of the experiments based on low wave vector magnons DMI is treated as a scaler.

\begin{table}[t!]
\caption{The DMI induced magnon stiffness constant D$_{DMI}$, measured for different systems. The abbreviation PPSWS stands for propagating spin wave spectroscopy \cite{Lee2015}. } \label{Tab:DMI_Linear}
\begin{center}
\begin{tabular}{l c c}
\hline  \hline
 System & D$_{DMI}$ [THz\AA] & Experimental methode\\
 \hline
 Fe/W(110) \cite{Zakeri2010, Zakeri2012a} & $2.9 \pm 0.5$ & SPEELS\\
 Pt/Co/MgO \cite{Lee2015} & 0.2 & PPSWS\\
 Ni$_{80}$Fe$_{20}$/Pt  \cite{Nembach2015} & 0.15 & BLS\\
 Pt/Co/AlO$_x$ \cite{Cho2015} & 1.25 & BLS\\
 Pt/Co/Ni \cite{Di2015} &0.5 & BLS \\
 \hline  \hline
\end{tabular}
\end{center}
\end{table}

In order to provide a comparison between the strength of DMI in different magnetic structures in Tab. \ref{Tab:DMI_Linear} we summarize the value of $D_{DMI}$ measured by different experimental methods. This quantity is simply determined by the linear relation between $\Delta \varepsilon$ (or $\Delta \omega$) and $Q$. The experimental values reported in literature are fitted by Eq. (\ref{Eq:DMBI_2}) and the value of $D_{DMI}$ is obtained. The largest value of $D_{DMI}$ is reported for the Fe bilayer on W(110), while the smallest one is found for the Ni$_{80}$Fe$_{20}$/Pt structure. It is important to notice that in the experiments based on low-wave vector excitations  the nonreciprocal contributions, caused by dipolar and anisotropy fields, are not usually subtracted. This may lead to a large uncertainty of $D_{DMI}$ obtained in this way.

\section{Summary} \label{Sec:Summary}

The pattern of the magnetic exchange parameters in a few atomic layers of a $3d$ ferromagnetic element grown on a nonmagnetic substrate can be very complex, even though the whole structure may show a rectangular hysteresis loop. Magnon spectroscopy provides a unique way of quantifying magnetic exchange interaction in such layered magnetic structures. The pattern of the exchange parameters can be measured by analyzing the full spectrum of magnon dispersion relation and comparing it with the results of first principles calculations. Of particular interest are the exchange parameters at the interface. These quantities can be quantified by probing the acoustic magnon mode of the system. The approach can be used for single element multilayers as well as multilayers composed of layers of different materials.

Tetragonally distorted ferromagnetic films may exhibit a large antiferromagnetic exchange interaction. The signature of this unusual antiferromagnetic exchange interaction can be observed in the magnon dispersion relation at the high-symmetry points.

In addition to the symmetric HEI the antisymmetric DMI can also be measured by probing the energy asymmetry of the dispersion relation of magnons propagating orthogonal to the sample magnetization. This is the most straightforward and direct way of probing DMI.

The magnon spectroscopy approach, described in this Topical Review, can be applied to a vast verity of magnetic films and multilayers grown on nonmagnetic substrates, in order to quantify the strength of the symmetric as well as the antisymmetric exchange interaction.

\section*{Acknowledgments}
The author would like to gratefully thank J. Kirschner for various discussions on the topic, strong support and continuous encouragement. Thanks to all the former members of the SPEELS team at the Max Planck Institute of Microstructure Physics for their contribution to the experiments discussed here. Fruitful collaborations and discussions with A. Ernst, L.M. Sandratskii and P. Buczek on the theory of magnetic excitations are gratefully acknowledged.
\section*{References}

\bibliographystyle{unsrt}
\bibliography{Zakeri_Refs}

\end{document}